\documentclass[twocolumn, english,aps, prb, amsmath]{revtex4}
\usepackage{newcent}

\usepackage[T1]{fontenc}
\usepackage[latin9]{inputenc}
\usepackage{color}
\usepackage{babel}

\usepackage{float}
\usepackage{amstext}
\usepackage{graphicx}
\usepackage{esint}
\usepackage[unicode=true, pdfusetitle,
 bookmarks=true,bookmarksnumbered=false,bookmarksopen=false,
 breaklinks=false,pdfborder={0 0 1},backref=false,colorlinks=false]
 {hyperref}

\makeatletter
\@ifundefined{textcolor}{}
{%
 \definecolor{BLACK}{gray}{0}
 \definecolor{WHITE}{gray}{1}
 \definecolor{RED}{rgb}{1,0,0}
 \definecolor{GREEN}{rgb}{0,1,0}
 \definecolor{BLUE}{rgb}{0,0,1}
 \definecolor{CYAN}{cmyk}{1,0,0,0}
 \definecolor{MAGENTA}{cmyk}{0,1,0,0}
 \definecolor{YELLOW}{cmyk}{0,0,1,0}
 }

\@ifundefined{showcaptionsetup}{}{%
 \PassOptionsToPackage{caption=false}{subfig}}
\usepackage{subfig}
\makeatother

\begin{document}

 \title{Dynamics of a magnetic dimer with exchange, dipolar and Dzyalozhinski-Moriya interaction}

\author{A. F. Franco, J.M. Martinez, J.L. Déjardin, and H. Kachkachi}

\affiliation{LAMPS, Université de Perpignan Via Domitia, 52 avenue Paul Alduy,
F-66860 Perpignan Cedex, France}
\begin{abstract}
We investigate the dynamics of a magnetic system consisting of two magnetic moments coupled by either exchange, dipole-dipole, or Dzyalozhinski-Moriya interaction. 
We compare the switching mechanisms and switching rates as induced by the three couplings. 
For each coupling and each configuration of the two anisotropy axes, we describe the switching modes and, using the kinetic theory of Langer, we provide (semi-)analytical expressions for the switching rate. We then compare the three interactions with regard to their efficiency in the reversal of the net magnetic moment of the dimer.
We also investigate how the energy barriers vary with the coupling. For the dipole-dipole interaction we find that the energy barrier may either increase or decrease with the coupling depending on whether the latter is weak or strong. 
Finally, upon comparing the various switching rates, we find that the dipole-dipole coupling leads to the slowest magnetic dimer, as far as the switching of its net magnetic moment is concerned.
\end{abstract}
\maketitle

\section{Introduction}

Multilayered magnetic systems as permanent magnets with high performances
encounter a renewed interest due to their potential use as magnetic
recording media with high thermal stability and reduced switching
fields \foreignlanguage{american}{\citep{knehaw91i3e,vicshe05i3e,suessetal05j3m}}.
Another candidate for information storing media, also of constantly growing
interest, is provided by magnetic nanoparticles. For such applications,
it is very important to understand how the dynamics of these systems
and their reversal mechanisms are altered by a change in their physical
parameters such as the underlying material, thickness, and stacking
conditions. There are also rather involved issues related with surface
and interface effects even if one assumes, to some extent, that the
problems of crystalline diffusion are properly dealt with during fabrication.
Apart from the issues related with the intrinsic properties of the
constituting elements (layers or particles), an issue of paramount
importance is that of (inter-layer or inter-particle) interactions
as these affect energy barriers which they have to circumvent during their switching process. Now, because the latter play a central
role in magnetic recording technology, the problem of interactions
must be addressed in a broader way. Indeed, several types of interactions
may occur in particle ensembles or in multi-layered systems. However,
it is a difficult task, if at all possible, to tell in detail what
interactions are involved in the dynamics of the systems to be studied.
In general, one resorts to the ``molecular field'' approach and
considers only the effective interaction. 
On the other hand, it is clear that the dynamics strongly depends on the type of interaction
that is operating within the system. As such, it would be useful to
relate the observed dynamics to a given type of interaction, even
as an effective one. In spin systems, there are short-range as well
as long-range, isotropic or anisotropic, interactions. These are mostly
exchange interactions (EI), dipole-dipole interactions (DDI), or Dzyalozhinski-Moriya
interactions (DMI).

On the atomic level, for instance, these three interactions are the most relevant spin-spin interactions and studying them in a systematic way is necessary in order to understand how a local spin excitation would propagate through the magnetic media as it is conveyed by each of these interactions. In particular, this is relevant in the pump-probe like experiments where one is interested in the long-range effect of a (local) demagnetization by a laser or any other source of local heating. 

Another example of application of the model considered in this work is that of two magnetic layers, each represented by its macroscopic magnetic moment, separated by a nonmagnetic spacer and coupled through the latter by an effective interaction which is either EI, DDI, or DMI. The static and dynamic properties of such multilayers with effective interactions have been studied experimentally and theoretically by many groups, see e.g., Refs. \onlinecite{CochranEtal_prb42, GrimsditchEtal_prb54, ZivieriEtal_prb62} where ultrathin multilayers were studied by the Brillouin Light Scattering technique and the effective coupling was estimated.
This model can also be adapted to a pair of macroscopic moments representing two nanoparticles, embedded in a nonmagnetic matrix. Furthermore, this somewhat toy problem may serve as a benchmark for many-particle systems. Moreover, for studying the dynamics of interacting systems one has to understand the dynamics of the elementary brick of two interacting elements and this can be represented as a magnetic dimer (MD).

One of the questions that we address here is how each of these interactions
affects the dynamics and, in particular, the magnetization reversal
of the MD. For this we first (semi)-analytically compute the switching
rates in several situations. Knowing the switching time a comparison
with \emph{e.g.} Network Analyzer-Ferromagnetic Resonance(NA-FMR) measurements, should
help us to tell which interaction is most relevant in the system studied. 

In the absence of thermal fluctuations, the switching is entirely
deterministic and occurs at some critical value of the effective field,
at which the (effective) energy barrier vanishes. In the case of DDI the
switching dynamics has been studied in the past by many authors, see
for instance Refs. \citep{bermal40jap,bermal41jap}. 
For classical systems, thermal effects on the dynamics of an MD can be accounted
for within the Langevin approach by solving the system of two coupled
Landau-Lifshitz equations, one for each magnetic moment, where the
effective deterministic field is augmented by the (random) Langevin
field \citep{lybcha93jap}. The equivalent approach that consists
in solving the corresponding Fokker-Planck equation (FPE) can also
be used \citep{rodeetal23ieee}. In the case of high-to-intermediate
damping regime and high-energy barriers, analytical expressions have
been obtained for the exchange-coupled MD \citep{kac03epl-kac04jml}
using Langer's approach \citep{lan68prl-lan69ap} and the results
have been favorably compared with the numerical approach based on
the solution of FPE \citep{titovetal05prb}. In Ref. \citep{kac03epl-kac04jml}
it was shown that there exists a critical exchange coupling below
which the MD reverses its direction via a two-step process, \emph{i.e.
}through\emph{ }a fanning mode, whereas above this critical coupling
the system switches through a coherent mode. At the critical coupling,
some saddle points become flat and Langer's approach, based on the
quadratic expansion of the energy, ceases to be valid. As a consequence,
the switching rate presents two disconnected branches corresponding
to the weak and strong coupling regimes. In Ref. \citep{titovetal05prb}
the FPE is transformed into an eigenvalue problem and the latter is
then solved using the numerical technique of matrix-continued fractions.
This work renders a smooth switching rate for exchange coupling strengths
that compares very well with the analytical asymptotes outside the
critical region. Finally, it is worth mentioning the work of Solomon
\citep{solomon99pr} where switching processes were studied and the
longitudinal and transverse switching times were computed for a quantum
MD at room temperature.

In the present work, we use Langer's approach to investigate the dynamics
of an MD with three different couplings, EI, DDI, and DMI.
The objective here is to compare the dynamics and switching modes
in each case with the aim to provide an answer as to which coupling is the most efficient
as far as the full switching of the MD is concerned and which configuration
is the most optimal for applications. To accomplish this, we consider two orientations
of the two anisotropy axes with respect to the bond axis, namely
the longitudinal (LA) and transverse (TA) anisotropies. These two
anisotropy configurations mimic the two cases of magnetic films with
two limiting thicknesses. In order to carry the full analytical calculation
of the various switching rates and provide (approximate) sensible
analytical expressions thereof, the applied magnetic field has been
ignored in this work. The other reason for this restriction is the
need to investigate the three couplings and compare the dynamical
behavior they entail without the influence of the applied magnetic
field. In a subsequent work we intend to include, within a numerical
approach, the latter and consider more general configurations of the
anisotropy axes.

The paper is organized as follows. In the next section we write
all the contributions to the MD energy, introduce our notation, and present
the method for computing the various switching rates. In the next
section, we deal with the three types of interactions. We first summarize
the previous results for an EI-MD and then deal with DDI and DMI,
successively. We treat the three cases of i) longitudinal anisotropy (LA)
where both anisotropy axes are parallel to the MD bond, ii) transverse
anisotropy (TA), with the anisotropies perpendicular to the bond direction,
and iii) mixed anisotropy (MA) with one anisotropy axis parallel and
the other perpendicular to the bond. In each case we investigate the various 
coupling regimes and compute the corresponding switching rates. The next section is devoted to a case study of a comparison between the three interactions. 
\section{Energy and switching rate}

In this section we define our notations and provide the basic formulae
of our calculations, including the energy and the switching rate.
For the latter we briefly summarize Langer's kinetic theory that will be used
in the case of intermediate-to-high damping (IHD) regime.

\subsection{Energy\label{sub:Energy}}

\begin{figure}[H]
\begin{centering}
\includegraphics[width=8cm]{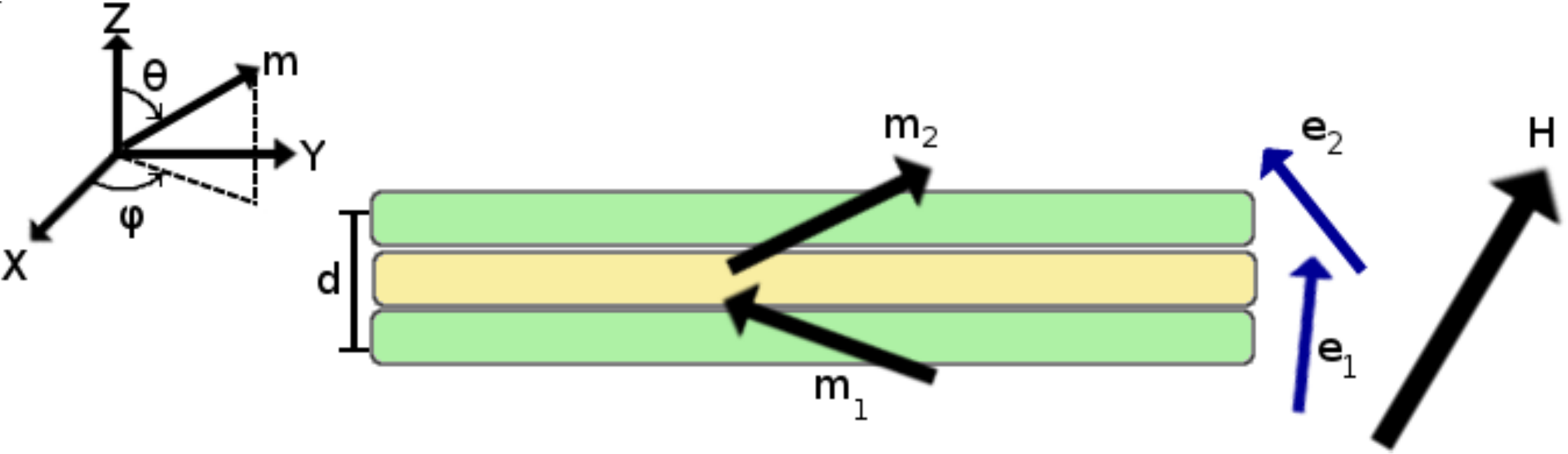} 
\par\end{centering}
\caption{\label{fig:MDSetup}Setup of the DDI-MD with oblique magnetic field
and arbitrary anisotropy axes.}
\end{figure}

As mentioned in the introduction, the system we study consists of
two macroscopic magnetic moments $\mathbf{m}_{i},i=1,2$, which may
represent two magnetic layers or two magnetic nanoparticles or still two atomic moments. 
Each magnetic moment has an effective uniaxial anisotropy that is assumed to result
from magneto-crystalline and/or shape anisotropy. In the case of multi-layered
systems, we consider both in-plane and out-of-plane anisotropy, thus
modeling magnetic films with two limiting thicknesses. The two layers
are coupled via the nonmagnetic layer by an effective EI, DDI, or DMI
and the corresponding dimer will then be referred to as
the EI-MD, DDI-MD, or DMI-MD, respectively. While exchange coupling
is invariant under global rotation of the system, DDI and DMI are
anisotropic as they originate from a coupling to the lattice. In particular,
the magnetic state induced by DDI depends on the orientation of the
vector connecting the magnetic moments, the MD bond. For the DDI-MD,
for definiteness, we set the latter in the $z$ direction and the
corresponding verse will be denoted $\mathbf{e}_{12}$, \emph{i.e.},
$\mathbf{e}_{12}=\mathbf{e}_{z}$. The connecting vector $\mathbf{r}_{12}$
is then written as $\mathbf{r}_{12}=d\mathbf{e}_{12}$ where $d$
is the distance between the centers of mass of the two layers. Since 
the magnetic layers are assumed to be much thinner than the nonmagnetic 
spacer, the distance $d$ is approximately the thickness of the latter. 
The anisotropy axes $\mathbf{e}_{i},i=1,2$ and the applied field with 
the verse $\mathbf{e}_{h}$ are \emph{a priory} in arbitrary 
directions {[}see Fig. \ref{fig:MDSetup}{]}.
%

In the sequel, we will use spherical coordinates for all vectors involved.
Hence, for the magnetic moments we write $\mathbf{m}_{i}=m_{i}\,\mathbf{s}_{i}$,
with $\left\Vert \mathbf{s}_{i}\right\Vert =1$ and $\mathbf{s}_{i}(\theta_{i},\varphi_{i}),i=1,2$.
The applied field is written $\mathbf{H}=H\,\mathbf{e}_{h}$, with
$\left\Vert \mathbf{e}_{h}\right\Vert =1$ and $\mathbf{e}_{h}(\theta_{h},\varphi_{h})$,
and the anisotropy axes are $\mathbf{e}_{i}(\theta_{i}^{\left(a\right)},\varphi_{i}^{\left(a\right)})$. $\theta$ and $\phi$ are respectively the polar and azimuthal angles as defined in Fig. \ref{fig:MDSetup}. Note also that the applied field $H$ is to be understood as $\mu_{0}H$ which is counted in $\mathrm{Tesla}$.

The energy of the MD then reads

\begin{eqnarray}
E & = & E_{\mathrm{Z}}+E_{\mathrm{A}}+E_{\mathrm{Int}}\label{eq:DM-Energy}
\end{eqnarray}
where $E_{\mathrm{Z}}$ is the Zeeman energy
\begin{equation}
E_{\mathrm{Z}}=-\mathbf{H}\cdot\sum_{i=1,2}\mathbf{m}_{i}=-\sum_{i=1,2}m_{i}\left(\mathbf{H}\cdot\mathbf{s}_{i}\right),\label{eq:DM-Z}
\end{equation}
$E_{\mathrm{A}}$ is the (uniaxial) anisotropy contribution
\begin{equation}
E_{\mathrm{A}}=-\sum_{i=1,2}K_{i}V_{i}\left(\mathbf{s}_{i}\cdot\mathbf{e}_{i}\right)^{2}.\label{eq:DM-An}
\end{equation}

The interaction energy $E_{\mathrm{Int}}$ may stem from the exchange (ferromagnetic) coupling
\begin{equation}
E_{\mathrm{Int}}=E_{\mathrm{Exch}}=-J\,\mathbf{s}_{1}\cdot\mathbf{s}_{2},\label{eq:DM-J}
\end{equation}
from the DM coupling
\begin{equation}
E_{\mathrm{Int}}=E_{\mathrm{DM}}=-\mathbf{D}\cdot\left(\mathbf{s}_{1}\times\mathbf{s}_{2}\right),\label{eq:DM-DM}
\end{equation}
or from the DDI contribution
\begin{equation}
E_{\mathrm{Int}}=E_{\mathrm{DDI}}=\left(\frac{\mu_{0}}{4\pi}\right)\left(\frac{m_{1}m_{2}}{d^{3}}\right)\mathbf{s}_{1}\cdot\mathbf{\mathcal{D}}_{12}\mathbf{s}_{2}\label{eq:DM-DDI}
\end{equation}
with
\begin{equation}
\mathbf{\mathcal{D}}_{12}\equiv3\left(\leftarrow\cdot\mathbf{e}_{z}\mathbf{e}_{z}\cdot\rightarrow\right)-1.\label{eq:DDI-tensor}
\end{equation}
being the DDI tensor. 

Let's recall that DMI is an anti-symmetrical
exchange interaction coming from a combination of low symmetry
and spin-orbit coupling \citep{dzyaloshinsky58jpcs,moriya60prl}.
In the presence of disorder, especially at the interface of thin films
or multilayers, the DMI has been shown to play an important role since
local symmetry is broken by surface effects. Indeed, it leads to large
anisotropy and may even change the magnetic order, see Ref. \citep{crelac98jmmm}
and references therein. In particular, it has been shown that DMI
is induced by spin-orbit coupling between two ferromagnetic layers
separated by a paramagnetic layer \citep{xiaetal97prb}. Accordingly,
in the present study, it is also relevant to investigate its effect
on the dynamics of the MD, on the same footing as the (symmetrical)
effective exchange coupling. 


Investigating the general situation with arbitrary orientations for
the easy axes is rather involved and can only be dealt with numerically.
This will be done in a subsequent work. In the present
work, we choose to focus on the qualitative behavior of the various
interactions and investigate how they affect the dynamics of the system.
For this purpose, we consider a situation that can be dealt with analytically,
thus allowing for a simpler analysis of the underlying physics. More
precisely, we assume equal magnitudes for the two magnetic moments
with equal anisotropies (in direction and magnitude), \emph{i.e.} $\mathbf{e}_{1}\parallel\mathbf{e}_{2}$
and $K_{1}=K_{2}$; no external magnetic field.

In the sequel, we will measure the energy in units of the anisotropy energy
and thus write 
\[
\mathcal{E}\equiv\frac{E}{k_{B}T}=\sigma\frac{E}{KV}\]
where 
\[
\sigma\equiv\frac{KV}{k_{\mathrm{B}}T}\]
is the reduced anisotropy energy and also the reduced energy barrier
in the non-interacting case. Therefore, the MD energy reads {[}the
Greek indexes run over $x,y,z$ while the Roman indexes run over $1,2${]}
\begin{eqnarray}
\mathcal{E} =\displaystyle{\frac{E}{k_{B}T}}&= & \sigma\left[-2h\sum_\alpha\, e_{h,\alpha}\sum_{i=1,2}s_{i,\alpha}\right.\nonumber\\
& &-\left.\sum_{i=1,2}\sum_{\alpha,\beta}e_{i,\alpha}e_{i,\beta}s_{i,\alpha}s_{i,\beta}\right]
+\mathcal{E}_{\mathrm{Int}}\label{eq:EnMatrix}
\end{eqnarray}
with
\begin{equation}
\mathcal{E}_{\mathrm{Int}}=-\sigma\sum_{\alpha,\beta}s_{1,\alpha}\left[j\,\delta_{\alpha\beta}+\delta\sum_{\gamma}\varepsilon^{\alpha\beta\gamma}e_{d,\gamma}-\xi\,\mathcal{D}_{12}^{\alpha\beta}\right]s_{2,\beta}.\label{eq:EnergyInt}
\end{equation}
$\varepsilon^{\alpha\beta\gamma}$ is the fully antisymmetric Levi-Civita
tensor of rank $3$ and $\mathbf{e}_{d}$ is the verse of $\mathbf{D}$.
We have introduced also the following (dimensionless) parameters

\begin{eqnarray}
h &\equiv& \frac{Hm}{2KV},\quad j\equiv\frac{J}{KV},\nonumber\\
\quad\delta&\equiv&\frac{D}{KV},\quad\xi\equiv\left(\frac{\mu_{0}}{4\pi}\right)\left(\frac{m^{2}/d^{3}}{KV}\right)\label{eq:DimParamsMat}
\end{eqnarray}
which imply that all energies are measured in units
of the anisotropy energy. For instance, $h$ is the usual ratio of
the magnetic field $H$ to the anisotropy field 

\begin{equation}
H_{A}=\frac{2KV}{m}=\frac{2K}{M_{s}}=\frac{2K_{a}}{\mu_{a}}.\label{eq:AnisotropyField}
\end{equation}

\subsection{Relaxation rate}

It was shown in Refs. \citep{kac03epl-kac04jml} that Langer's expression
\citep{lan68prl-lan69ap} for the escape rate from the minimum $(\theta^{(m)},\varphi^{(m)})$
through the saddle point $(\theta^{(s)},\varphi^{(s)})$ takes the
more compact form

\begin{equation}
\Gamma=\frac{\left|\kappa\right|}{2\pi}\frac{\tilde{Z}_{s}}{Z_{m}},\label{eq:LangerRR}
\end{equation}
where $Z_{m}$ and $Z_{s}$ are respectively the partition functions
computed in the vicinity of the minimum and the saddle point and $\left|\kappa\right|$ is the
attempt frequency. The latter represents the growth rate of a nucleating fluctuation at the
saddle point and thus characterizes the unstable barrier-crossing mode.
This expression indicates that the escape rate is simply given by the ratio of the total current
through the saddle point to the number of particles (or points in the system phase space) in the
metastable state.
In fact, within Langer's approach the problem of calculating the switching rate for a
multi-dimensional process is reduced to solving a steady-state Fokker-Planck equation for the
probability density $\rho$, \emph{i.e.}, $\partial_t \rho=0$, in the immediate neighborhood of the
saddle point that the system crosses as it goes from a metastable state to another state of greater
stability.
The probability density $\rho$ is connected to the probability current via the continuity
equation. On the other hand,  $\rho$ can be written as $\rho_\text{eq}=e^{-\beta\mathcal{H}}/Z$
times some other function. From these two relations one can write the probability
current in terms of the partition function \cite{braun_jap94, garaninEtal_pre99}, see also Ref.
\onlinecite{coffeyetal01acp} for great details.
Now, since the switching rate is given by the total probability flux through a surface near the
saddle point, Langer's result for the escape rate can be achieved by computing the energy-Hessian
eigenvalues near the saddle points and metastable states. From the latter, one then infers the
partition function $\tilde{Z_{s}}$ of the system restricted to the region around the saddle point
and the partition function $Z_{m}$ of the region around the metastable state. When computing these
partition functions, one has to identify and take care of each Goldstone mode, that is a massless
mode or zero-energy fluctuation associated with a continuous unbroken global symmetry. 
The tilde on $Z_{s}$ reminds us of the fact that the negative eigenvalue of the energy Hessian corresponding to the escape route is (formally) taken with the absolute value \footnote{See Ref.~\cite{lan68prl-lan69ap} for a rigorous derivation}. $Z_{s}$ is the product of contributions from all eigenvalues. 

In Langer's approach the attempt frequency $\kappa$ is computed by
linearizing the Landau-Lifshitz equation around the saddle point,
diagonalizing the resulting transition matrix\cite{coffeyetal01acp,kac04jml},
and selecting its negative eigenvalue. However, only in a few situations
can $\kappa$ be obtained analytically. In fact, in the general situation,
$\kappa$ can only be computed numerically. Accordingly, one computes
the unique
\footnote{Indeed, if the saddle point is to describe the nucleating fluctuation,
there must be exactly one direction of motion away from the saddle
point in which the solution of the equations of motion of the modes
$\psi_{n}$ is unstable \cite{lan68prl-lan69ap}.
} negative eigenvalue $\kappa$ of the steady-state FPE corresponding
to the unstable mode at the saddle point as the negative eigenvalue
of the dynamic matrix $\tilde{M}_{mn}=-\lambda_{n}(PMP^{T})_{mn}$,
where $\lambda_{n}$ are the eigenvalues of the energy Hessian at the saddle point, $M$ the dynamic matrix, and $P$ is the transformation matrix from the initial coordinates to the {}``canonical'' ones.

Therefore, for a given elementary process, \emph{i.e.}, an escape
from the minimum $(\theta^{(m)},\varphi^{(m)})$ through the saddle
point $(\theta^{(s)},\varphi^{(s)})$, we have to compute the partition
function
\[
Z=\int\left(\mathcal{D}\mathbf{s}\right)\, e^{-\beta E\left(\mathbf{s}\right)}\]
at the saddle and metastable states. For this, we perform
a quadratic expansion of the energy at these stationary states. This
is where Langer's approach meets its limit of validity because such an expansion
is only meaningful when the stationary point is well defined. More precisely,
Langer's approach is only valid in the case of high energy
barriers $\Delta E$, \emph{i.e.}, when $\beta\Delta E\gg1$ and also
intermediate-to-high damping\cite{garaninEtal_pre99,coffeyetal01acp}. 

In the case of a two-body problem, such as that of MD, in the weak
coupling regime the magnetization of the whole system switches in
a two-step process; an example is shown in Fig. \ref{fig:EI-MD} in the case of EI-MD.
The first step of switching corresponds to the passage of the first
magnetic moment from the initial state into an intermediate state
through the saddle point. This step lasts the (switching) time $\tau_{1}$.
The second step is taken by the second magnetic moment that then proceeds
to switch through a second saddle point and this step lasts the time
$\tau_{2}$. The total time required by the MD to switch is then $\tau=\tau_{1}+\tau_{2}$,
and in terms of the switching rate ($\Gamma=\tau^{-1}$), one has

\begin{equation*}
\frac{1}{\Gamma}=\frac{1}{\Gamma_{1}}+\frac{1}{\Gamma_{2}}. 
\end{equation*}

Next, for fully identical magnetic moments one has to consider the left-right symmetry and multiply the expression above by a factor of $2$ leading to the final expression
for the switching rate corresponding to the two-step process of the MD

\begin{equation}
\Gamma_{\mathrm{total}}=2\frac{\Gamma_{1}\Gamma_{2}}{\Gamma_{1}+\Gamma_{2}}.\label{eq:Two-StepProc-RR}
\end{equation}

In fact, one may have other symmetry factors depending on the system setup.

Consequently, in the sequel our task will consist in analyzing the
energy potential surface in each situation, studying the various switching paths,
and combining the corresponding switching rates according to Eq.
(\ref{eq:Two-StepProc-RR}). For each elementary step corresponding
to an escape from a minimum through a saddle point we will use Eq.
(\ref{eq:LangerRR}) to compute the corresponding switching rate.

Defining the characteristic time of the underlying material
$t_{s}=\left(\gamma H_{A}\right)^{-1}=\mu_{a}/(2\gamma K_{a})$,
where $\gamma\simeq1.76\times10^{11}$ (T.s)$^{-1}$ is the gyromagnetic
factor, the final (dimensionless) switching rate may be given in
$s^{-1}$ upon multiplying by

\begin{equation}
\frac{\gamma k_{B}T}{\mu_{a}}=\frac{1}{2}\left(\frac{2\gamma KV}{\mu_{s}}\right)\left(\frac{k_{B}T}{KV}\right)=\frac{1}{2}\frac{t_{s}^{-1}}{\sigma}.\label{eq:GammaScalingFactor}
\end{equation}

For cobalt, for instance, we have $ \mu_a=1.57\times 10^{-23} A m^{-1} atom^{-1}$, $K_a = 2.53\times 10^{-24} J atom^{-1}$, leading to
$t_s=1.76\times10^{-11} s$.

In this work we compute the relaxation rate by combining Langer's
approach, which is valid in the IHD regime, and the Landau-Lifshitz
equation (LLE) with its phenomenological damping parameter. Using
the LLE for obtaining the attempt frequency in the prefactor in Eq.
(\ref{eq:LangerRR}) leads to spurious effects when one formally
takes the limits $\alpha\rightarrow0$ or $\alpha\rightarrow\infty$.
\footnote{We thank Yu. Kalmykov for having reminded us to mention this issue.
We also thank D. Garanin and O. Chubykalo-Fesenko for a discussion
of related issues.%
} However, Langer's expression for the relaxation rate may be ``regularised''
by using Gilbert's damping instead of the Landau-Lifshitz damping.
Indeed, the latter may be shown to be identical to Gilbert's if the
gyromagnetic factor $\gamma$ is replaced by $\gamma^{*}=\gamma/\left(1+\alpha^{2}\right)$.
In the present calculations this amounts to replacing the scaling
time $t_{s}$ defined above by $t_{s}^{*}=t_{s}\left(1+\alpha^{2}\right)$.
In fact, Landau-Lifshitz and Gilbert's equations are related by the
transformation
\[
\gamma\rightarrow\frac{\gamma}{1+\alpha^{2}},\qquad\alpha\rightarrow\frac{\alpha}{1+\alpha^{2}}.
\]

We recall that our main objectives in this work are: i) an investigation
of the behaviour of the relaxation rate as a function of the MD coupling
and ii) a pairwise comparison of three types of layer coupling (exchange,
dipolar, and Dzyaloshiski-Moriya). In particular, we do not investigate
the damping dependence of the relaxation rate. A thorough study of
all crossovers between the various damping regimes is given in Ref.
\onlinecite{garaninEtal_pre99} where one can see that the boundaries
between the regimes are not simply $\alpha=1$.

\section{Exchange coupled magnetic dimer}

\begin{figure*}
\includegraphics[scale=0.8]{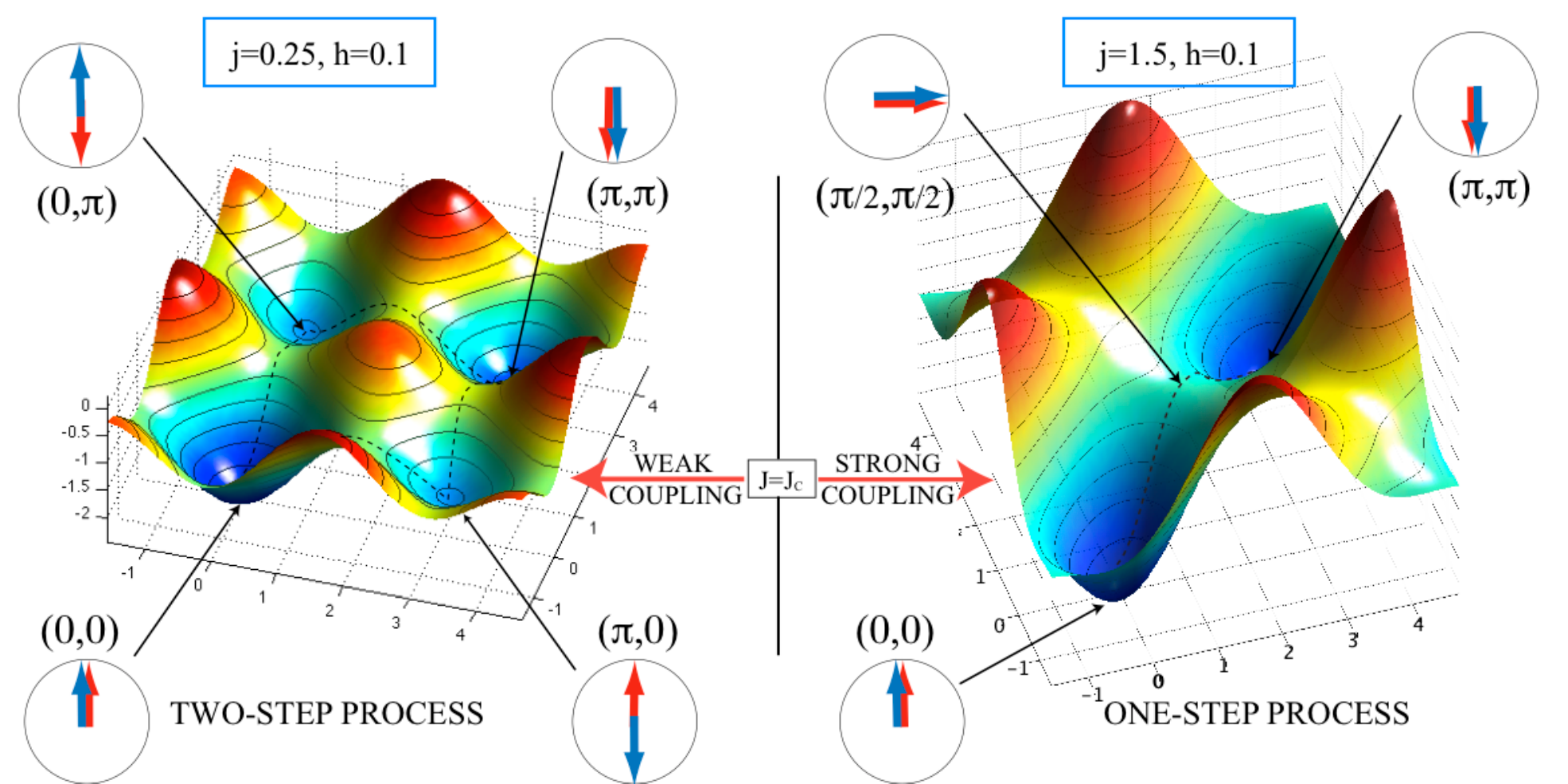}
\caption{\label{fig:EI-MD}Crossover from a two-step to a one-step switching
to its stable state (SS) of an EI-MD with LA. }
\end{figure*}

In this section, for later use, we briefly summarize the results of Ref. \citep{kac03epl-kac04jml}.
The situation is sketched in Fig. \ref{fig:EI-MD}.

In the case of parallel easy axes and longitudinal field, it was found
that there is a critical exchange coupling $j_{c}$ that depends on
the applied field and anisotropy constant, \emph{i.e.} $j_{c}(H,K)$,
above which the MD behaves as a macrospin with a double energy barrier
that switches from a metastable state to a more stable one in a coherent
manner, see Fig. \ref{fig:EI-MD} (right). 
Below $j_{c}$ the system is weakly coupled and switches in a two-step process through two different escape routes (saddle points), see Fig. \ref{fig:EI-MD} (left). Setting $\mathbf{e}_{h}\parallel\mathbf{e}_{1}\parallel\mathbf{e}_{2}$, in the notations of Eq. (\ref{eq:DimParamsMat}), the critical exchange coupling $j_{c}$ was found to be 

\begin{equation}
j_{c}=1-h^{2}.\label{eq:CritExchange}
\end{equation}

For $j>j_{c}$, the MD switches from the metastable state $\left(\pi,\pi,\varphi\right)$,
that is a ferromagnet, against the field, into the ferromagnetic state
$\left(0,0,\varphi\right)$, through the saddle point $\left(\arccos\left(-h\right),\arccos\left(-h\right),\varphi\right)$.
The angle $\varphi$ is arbitrary because of the uniaxial symmetry
and the equality of the spin polar angles is due to the fact that
the two spins are identical (same amplitude and same anisotropy).
The corresponding switching rate is given by (in zero field) 

\begin{eqnarray}
\Gamma_{j>j_{c}} & = & \alpha\sqrt{\frac{2\sigma}{\pi}}\frac{1+1/j}{\sqrt{1-1/j}}e^{-\sigma}.\label{eq:Gamma_tsp_j>jc}
\end{eqnarray}

In the weak-coupling regime $j<j_{c}$, the switching rate is obtained
by combining the switching rates corresponding to the two escape
routes taken by the two spins, see Fig. \ref{fig:EI-MD} (left). The result is somewhat more involved
and given in Refs. \cite{kac03epl-kac04jml}, see also Ref. \citep{titovetal05prb}.

The escape rate for the EI-coupled MD will be compared to the other two cases of DDI- and
DMI-coupled MD. For this purpose, we recall here the energy barriers (in the absence of the
magnetic field)
\begin{equation}
\Delta\mathcal{E} = \frac{\sigma}{2}(1\pm j).\label{eq:BarriersExchange}
\end{equation}
\section{Dipolar coupled magnetic dimer}

We consider both the longitudinal and transverse anisotropies, \emph{i.e.}
$\mathbf{e}_{1}\parallel\mathbf{e}_{2}\parallel\mathbf{e}_{12}\parallel\mathbf{e}_{z}$
and $\mathbf{e}_{1}\parallel\mathbf{e}_{2}\perp\mathbf{e}_{12}\parallel\mathbf{e}_{z}$,
respectively, which will be referred to as the LA and TA setup, respectively. Furthermore,
we will also discuss the case of mixed anisotropy, $\mathbf{e}_{1}\parallel\mathbf{e}_{12},\mathbf{e}_{2}\perp\mathbf{e}_{12}$,
which would mimic the case of an MD with a sufficiently thin film coupled
to a sufficiently thick one. Comparison with the exchange-coupled
MD will be done only in the case of LA, considered in Refs. \citep{kac03epl-kac04jml}.

A few general assumptions allow us to simplify the problem without
any loss of generality as far as the underlying physics is concerned.
Indeed, in the sequel, we will assume the following. If there is no magnetic field, 
the equilibrium orientation of the net magnetic moment is in the plane defined by the DDI axis and the anisotropy axes. The
latter are assumed to lie in the $xz$ plane, \emph{i.e.}, $\varphi_{1}^{\left(a\right)}=\varphi_{2}^{\left(a\right)}=\text{0}$.
Consequently, the energy becomes

\begin{eqnarray}
\mathcal{E} & = & -\sigma\sum_{i=1,2}\cos^{2}\left(\theta_{i}-\theta_{i}^{\left(a\right)}\right)\\
& & -\sigma\xi\left[2\cos\theta_{1}\cos\theta_{2}-\sin\theta_{1}\sin\theta_{2}\cos\left(\varphi_{1}-\varphi_{2}\right)\right].\nonumber
\label{eq:ZerofieldCase}
\end{eqnarray}

The analytical study will be further restricted to the following three
cases: 

\begin{enumerate}
\item Longitudinal anisotropy (LA): both anisotropy axes are parallel to
the MD axis $\mathbf{e}_{12}$, \emph{i.e.} $\theta_{1}^{\left(a\right)}=0=\theta_{2}^{\left(a\right)}$.
Moreover, due to the fact that all contributing fields are acting
in the same plane, the two magnetic moments of the MD will move in
the same plane so that $\varphi_{1}=\varphi_{2}$. 
\item Transverse anisotropy (TA): both anisotropy axes are perpendicular
to the MD axis, $\theta_{1}^{\left(a\right)}=\frac{\pi}{2}=\theta_{2}^{\left(a\right)}$. 
\item Mixed anisotropy (MA): one anisotropy axis is parallel to the MD axis
and the other perpendicular to it, $\theta_{1}^{\left(a\right)}=0,\theta_{2}^{\left(a\right)}=\frac{\pi}{2}$. 
\end{enumerate}

\begin{figure*}
\subfloat[$\xi=0.2$]{\includegraphics[width=8cm]{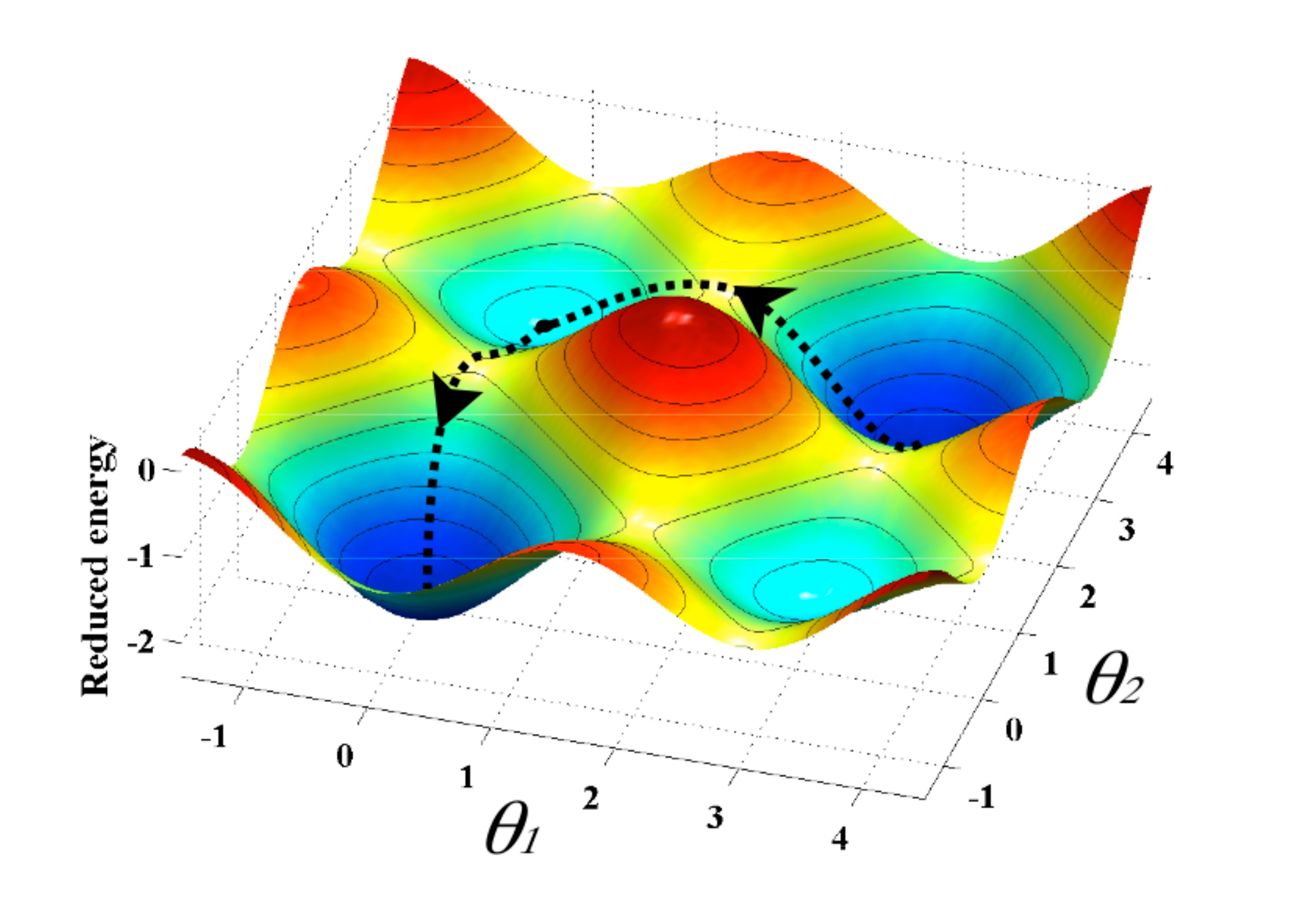}}
\subfloat[$\xi=0.4$]{\includegraphics[width=8cm]{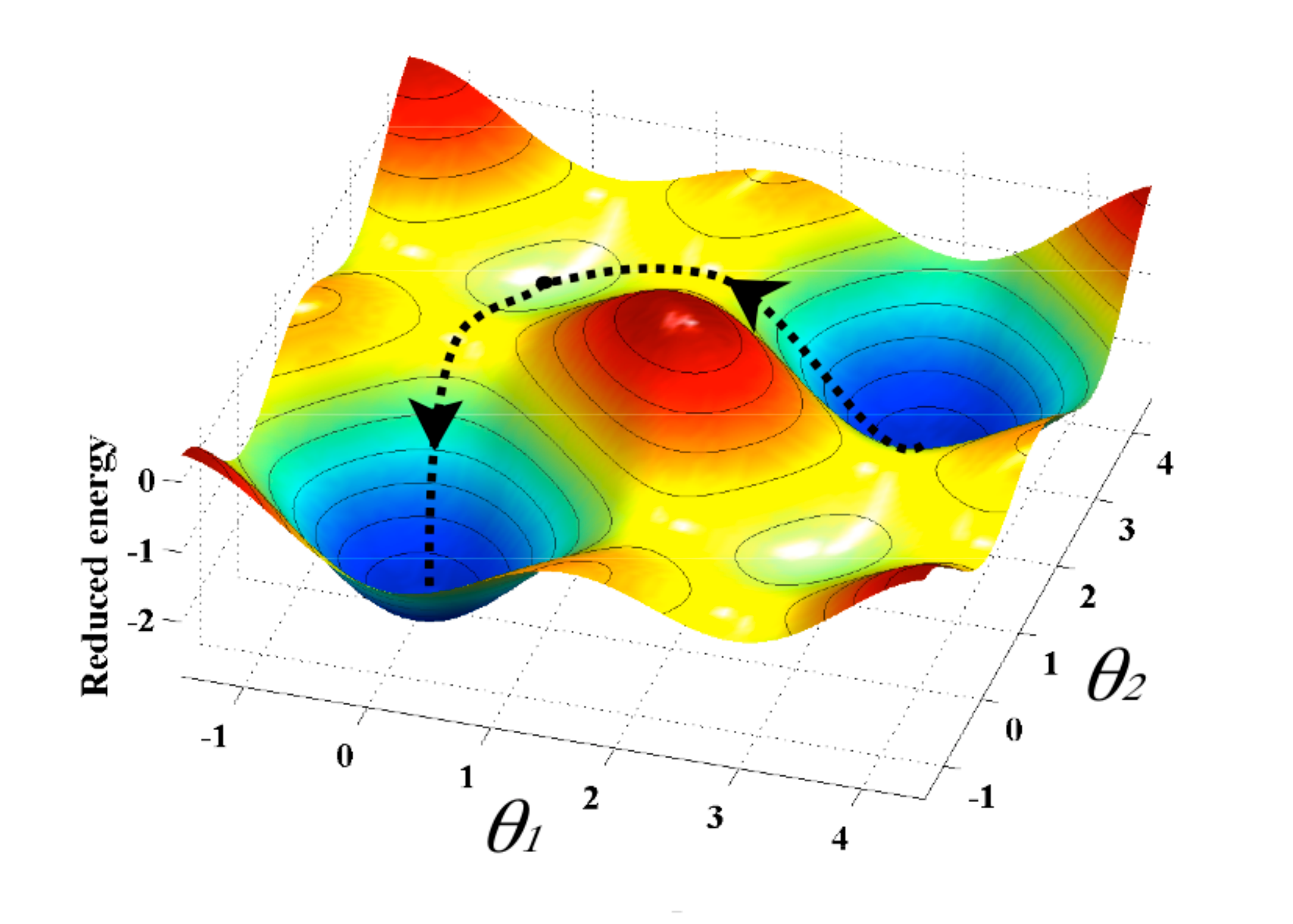}}\\
\subfloat[$\xi=0.6$]{\includegraphics[width=8cm]{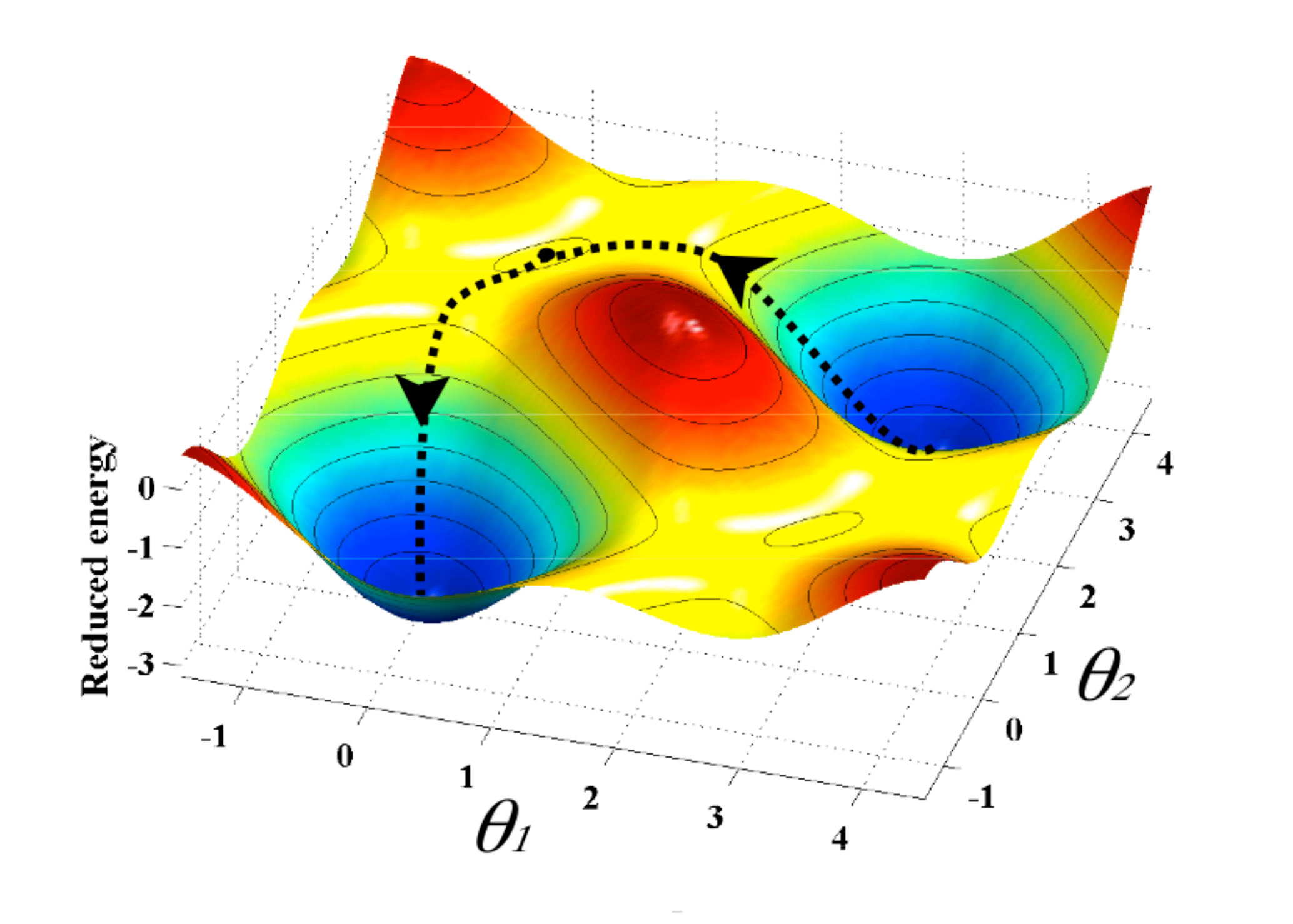}}
\subfloat[$\xi=1$]{\includegraphics[width=8cm]{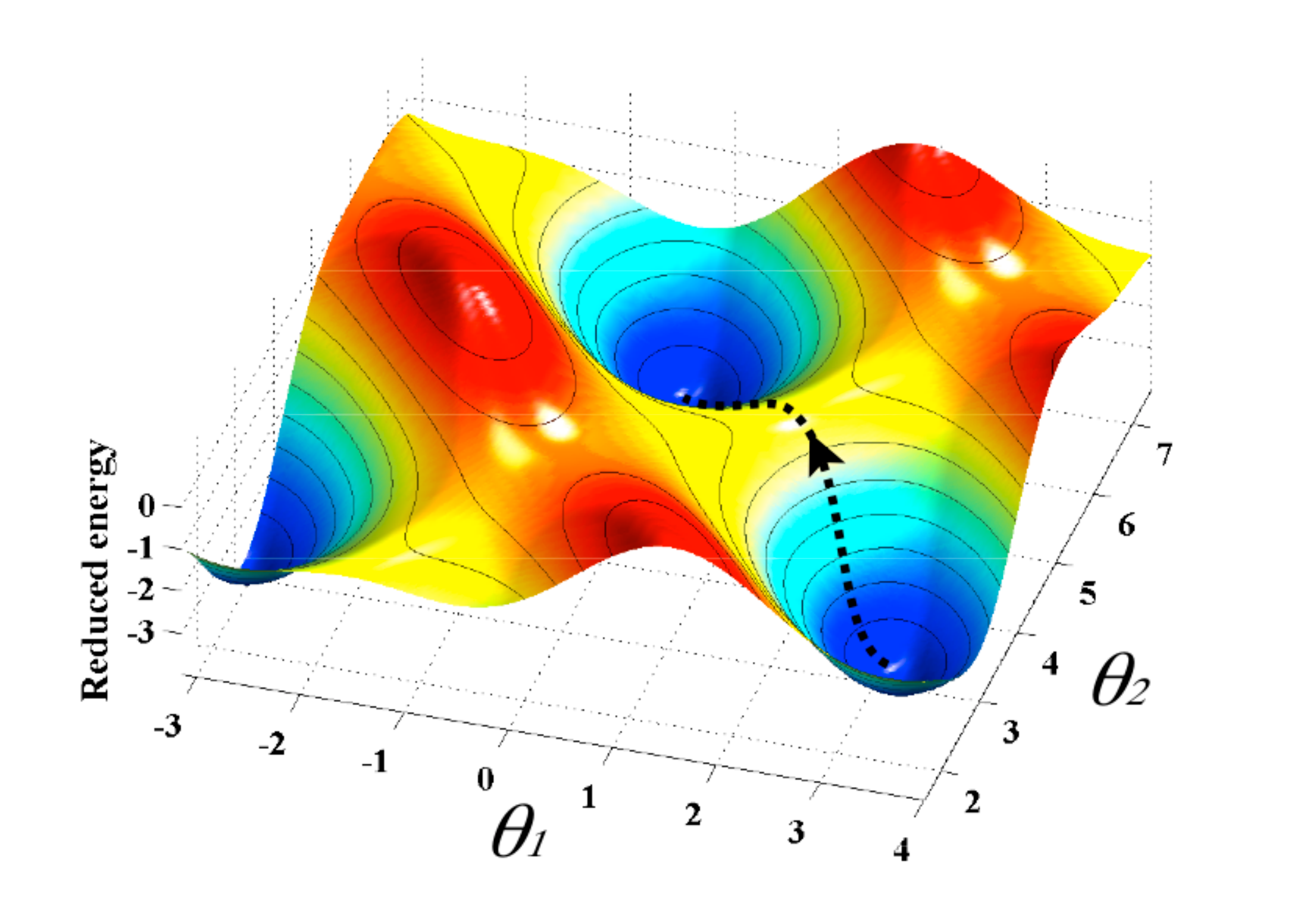}}
\caption{\label{fig:DDI-MD_LA}Evolution of the energy potential surface of a DDI-MD with longitudinal anisotropy configuration as $\xi$ increases, with $\sigma=1.5$. }
\end{figure*}

Differentiating with respect to the remaining variables,\emph{ i.e.}
the two polar angles $\theta_{i},i=1,2$, leads to the various equations
for the stationary states whose solutions depend on the anisotropy
setup.

\subsection{Longitudinal anisotropy}
The whole set of stationary states is given by
\begin{eqnarray*}
\left(\theta_{1},\theta_{2}\right) & = & \left(0,\pm\pi\right),\left(\pm\pi,0\right),\left(\pm\pi,\pm\pi\right),\left(\pm\pi,\mp\pi\right),\\
 &  & \left(\pm\frac{\pi}{2},\pm\frac{\pi}{2}\right),\left(\pm\frac{\pi}{2},\mp\frac{\pi}{2}\right)
\end{eqnarray*}
together with the following ones (which are saddle points) for $\xi\leq\frac{2}{3}$
or $\xi\geq2$,
\begin{equation}
\left(\cos\theta_{1},\cos\theta_{2}\right)=\left(\pm x_{1}^{\epsilon},\pm x_{2}^{\epsilon}\right),\left(\pm x_{1}^{\epsilon},\mp x_{2}^{\epsilon}\right)\label{eq: SP-LAWC}
\end{equation}
where
\begin{eqnarray}
x_{1}^{\epsilon} & = & \frac{1}{\sqrt{2}}\sqrt{1+\frac{3}{4}\xi^{2}-a^{\epsilon}},\qquad x_{2}^{\pm}=x_{1}^{\mp}\nonumber \\
a^{\epsilon} & \equiv & \epsilon\sqrt{\left(\frac{\xi^{2}}{4}-1\right)\left(9\frac{\xi^{2}}{4}-1\right)}\equiv\epsilon a,\quad\epsilon=\pm.\label{eq:Roots-Const_a}
\end{eqnarray}

It can be checked that the radicant is always positive and no additional
special ranges are found regarding the existence of the roots. However,
as $x_{1}^{\epsilon}$ and $x_{2}^{\epsilon}$ are cosines they must
satisfy $-1\leq x_{i}^{\epsilon}\leq1$, which is only true for $\xi\leq\frac{2}{3}$.
Hence we can identify two different regimes, the weak-coupling (WC)
regime $\xi\leq\frac{2}{3}$, and the strong-coupling (SC) regime
$\xi\geq\frac{2}{3}$. 
We will see later that this critical value corresponds to the vanishing
of the smallest eigenvalue of the energy Hessian at one of the energy
minima. It also marks the nucleation of a particular switching mode
and allows us to determine the nucleation field \citep{bermal40jap,bermal41jap}.

The energy potential surface for this situation is shown in Fig. \ref{fig:DDI-MD_LA} for a varying DDI strength $\xi$.

\subsubsection{Weak coupling ($\xi\leq\frac{2}{3}$)}
The minima correspond to ferromagnetic (FM) states along the DDI axis 
\begin{equation}
\left(\theta_{1},\theta_{2}\right)=\left(0,0\right),\left(\pi,\pi\right).\label{eq:Min-LAWC}
\end{equation}

The metastable states are the anti-ferromagnetic (AFM) states along the DDI axis
\begin{equation}
\left(\theta_{1},\theta_{2}\right)=\left(0,\pi\right),\left(\pi,0\right)\label{eq:MS-LAWC}
\end{equation}
and the maxima are the (anti)ferromagnetic states perpendicular to the DDI axis 
\begin{equation}
\left(\theta_{1},\theta_{2}\right)=\left(\pm\frac{\pi}{2},\pm\frac{\pi}{2}\right),\left(\pm\frac{\pi}{2},\mp\frac{\pi}{2}\right).\label{eq:Max-LAWC}
\end{equation}

Finally, the saddle points are located at
\begin{equation}
\left(\theta_{1},\theta_{2}\right)=\left(\varepsilon_{1}\arccos\left(\varepsilon_{2}x_{1}^{\gamma}\right),\varepsilon_{1}\arccos\left(\varepsilon_{2}x_{2}^{\gamma}\right)\right)\label{eq:SP-LAWC}
\end{equation}
where the signs $\varepsilon_{1}=\pm$ and $\varepsilon_{2}=\mp$ are independent of each other.

\begin{figure*}
 \includegraphics[width=16cm]{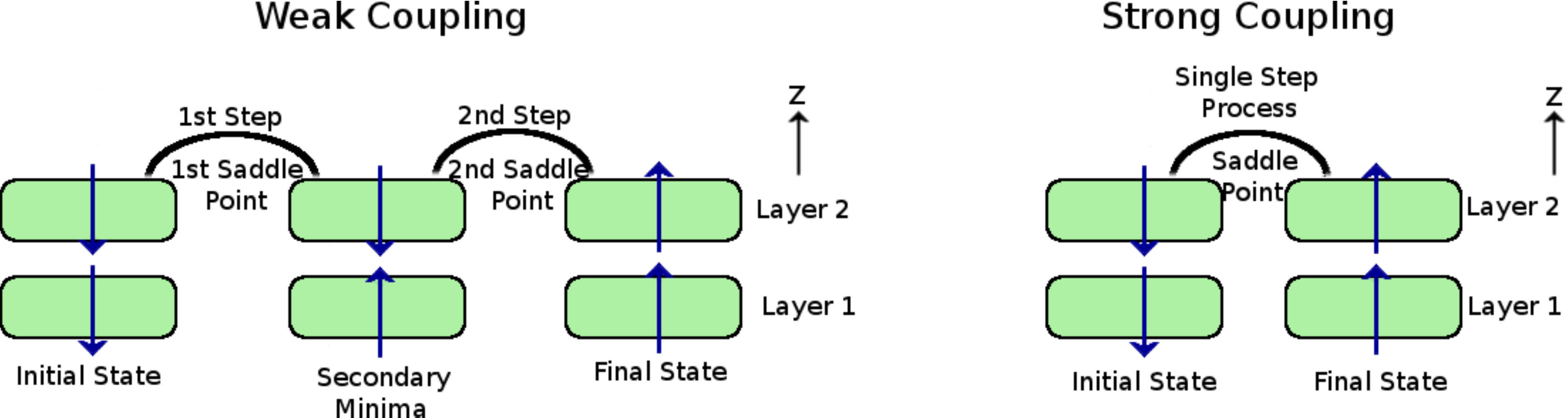}
 \caption{\label{fig:EscapeRoute-LAWCSC}Escape route of the MD with LA.}
\end{figure*}

Now we compute the switching rate in this coupling regime. In Fig. \ref{fig:DDI-MD_LA}a we see that the system goes through 
the following steps: i) from the state $\left(\pi,\pi\right)$ to the state $\left(0,\pi\right)$ through the saddle point 

\begin{equation}
\left(\cos\theta_{1}^{\left(1\right)}=x_{1}^{+},\cos\theta_{2}^{\left(1\right)}=-x_{2}^{+}\right)\label{eq:SP1-LAWC}
\end{equation}
{[}see Eq. (\ref{eq:SP-LAWC}){]}, and then ii) it passes from the
state $\left(0,\pi\right)$ to the state $\left(0,0\right)$ through the saddle point
\begin{equation}
\left(\cos\theta_{1}^{\left(2\right)}=x_{1}^{-},\cos\theta_{2}^{\left(2\right)}=-x_{2}^{-}\right).\label{eq:SP2-LAWC}
\end{equation}
These transitions are sketched in Fig. \ref{fig:EscapeRoute-LAWCSC} (left).

Therefore, in order to compute the switching rate corresponding to
the two-step process $\left(\pi,\pi\right)\rightarrow\left(0,\pi\right)\rightarrow\left(0,0\right)$
we need to compute the switching rate of each step and combine them according to the rule in Eq. (\ref{eq:Two-StepProc-RR}) 
where the individual switching rates are then computed using Langer's expression (\ref{eq:LangerRR}).

In order to compute the switching rate $\Gamma_{\left(\pi,\pi\right)\rightarrow\left(0,\pi\right)}$ we first compute the partition function at the minimum $\left(\pi,\pi\right)$ and at the saddle point $\left(\theta_{1}^{\left(1\right)},\theta_{2}^{\left(1\right)}\right)$, namely
\begin{equation}
Z_{\left(\pi,\pi\right)}\simeq\frac{\left(2\pi\right)^{2}}{\sigma^{2}\left(2+3\xi\right)\left(2+\xi\right)} e^{2\sigma\left(1+\xi\right)}.\label{eq:PF-Min1-LAWC}
\end{equation}

Likewise, the partition function at the
metastable state $\left(0,\pi\right)$ is

\begin{equation}
Z_{\left(0,\pi\right)}=\frac{\left(2\pi\right)^{2}}{\sigma^{2}\left(2-\xi\right)\left(2-3\xi\right)}e^{2\sigma\left(1-\xi\right)}.\label{eq:PF-Min2-LAWC}
\end{equation}

This result can also be found upon noting that since the anisotropy
is uniaxial, in order to change the system minimum from the ferromagnetic
state $\left(\pi,\pi\right)$ to the anti-ferromagnetic state $\left(0,\pi\right)$,
we can simply change the sign of the interaction, \emph{i.e}. replace
$\xi$ by $-\xi$. 

The lowest eigenvalue of the energy Hessian at the metastable minimum
is $\lambda=\sigma\left(2-3\xi\right)$. As mentioned earlier, we
see that when this eigenvalue vanishes it yields the critical value
for the DDI coupling, namely $\xi=2/3$. Indeed, the nucleation field
in this case is $h_{n}\propto\sigma\left(2-3\xi\right)$ which
coincides with the result of Refs. \onlinecite{bermal40jap,bermal41jap,rodeetal23ieee}. In Ref. \onlinecite{rodeetal23ieee} the authors
defined the parameter for the DDI strength $k_{\mathrm{int}}$ as
the ratio of the DDI field to the anisotropy field. In our notations,
$k_{\mathrm{int}}=\xi/2$. The critical value of $k_{\mathrm{int}}=1/3$
coincides with our condition $\xi=2/3$.

At the saddle point (\ref{eq:SP1-LAWC}) the energy is
\begin{eqnarray}
\mathcal{E}_{0}^{\left(1\right)} & = & \sigma\left(\frac{3}{4}\xi^{2}-1\right)\label{eq:En-SP1-LAWC}
\end{eqnarray}
and the energy barrier separating the minimum $\left(\pi,\pi\right)$
from the saddle point $\left(\theta_{1}^{\left(1\right)},\theta_{2}^{\left(1\right)}\right)$ is 

\begin{equation}
\Delta\mathcal{E}^{\left(1\right)}=\sigma\left(1+2\xi+\frac{3}{4}\xi^{2}\right).\label{eq:EnBar1-LAWC}
\end{equation}

This is plotted in Fig. \ref{fig:Energy-barrier} where we see that the (weak) DDI brings a
correction $\sigma\left(2\xi+\frac{3}{4}\xi^{2}\right)$ to the free MD energy barrier
$\Delta\varepsilon=\sigma$.
For ferromagnetic order ($\xi>0$), this correction enhances the energy
barrier and this is compatible with the fact that due to the ferromagnetic
coupling, it is more difficult for the first spin to switch.

Within Langer's approach to the calculation of the switching rate
the saddle point may retain a subgroup of the symmetry group of the
system, in which case some of the Hessian eigenvalues at the saddle
point vanish, and then a special treatment is required for this situation.
For the one-spin problem with uniaxial anisotropy, for instance, the saddle point has a $U(1)$
symmetry around the $z$ axis, or with respect to the rotation $\mathcal{R}\left(\mathbf{e}_{z},\varphi\right)$.
This leads to a vanishing eigenvalue of the Hessian at the saddle
point corresponding to fluctuations with respect to the angle $\varphi$
($\simeq\varphi^{s}+p$), where $\varphi^{s}$ is the value of the
azimuthal angle at the saddle point. Likewise, the saddle points (\ref{eq:SP1-LAWC},
\ref{eq:SP2-LAWC}) have rotational symmetry with respect to the azimuthal
angle and as such one should use the energy

\begin{eqnarray*}
\mathcal{E} & = & -\sigma\left[\cos^{2}\theta_{1}+\cos^{2}\theta_{2}\right]\\
 &  & -\sigma\xi\left[2\cos\theta_{1}\cos\theta_{2}-\sin\theta_{1}\sin\theta_{2}\cos\varphi\right]
\end{eqnarray*}
where $\varphi\equiv\varphi_{1}-\varphi_{2}$. The saddle points (\ref{eq:SP1-LAWC},
\ref{eq:SP2-LAWC}) should refer both to the polar angles $\theta_{i}$
and to a given value of the azimuthal angle $\varphi$, even if it is arbitrary.

Let us then perform the expansion around the saddle point $\left( \theta^{\left( s \right)}_{i},\varphi^{\left( s \right)}_{i} \right) $
\begin{equation*}
t_{i}=\theta_{i}-\theta_{i}^{(s)},p_{i}=\varphi_{i}-\varphi_{i}^{(s)},\, i=1,2
\end{equation*}
and expand the energy above to second order in the small variables
$t_{i},p_{i}$. Doing so we obtain
\begin{eqnarray*}
\mathcal{E}^{\left(1\right)} & \simeq & \mathcal{E}_{0}^{\left(1\right)}+2\sigma\left(t_{1},t_{2}\right)\left(\begin{array}{cc}
a & 0\\
0 & -a\end{array}\right)\left(\begin{array}{c}
t_{1}\\
t_{2}\end{array}\right)\\
 &  & +\frac{\sigma}{2}\left(\pi_{+},\pi_{-}\right)\left(\begin{array}{cc}
0 & 0\\
0 & \xi^{2}\end{array}\right)\left(\begin{array}{c}
\pi_{+}\\
\pi_{-}\end{array}\right)
\end{eqnarray*}
where $\displaystyle{\pi_{\pm}=\frac{p_{1}\pm p_{2}}{\sqrt{2}}}$.

Note that the mode corresponding to $\pi_{+}$ is a soft mode, \emph{i.e.},
a zero-energy mode, or still a Goldstone mode. This corresponds to
the $U\left(1\right)$ symmetry mentioned earlier and that must be dealt
with properly in order to avoid the divergence of the partition function.
In fact, integration over this variable simply yields the factor $2\pi$.
The partition function is then given by

\begin{equation}
\tilde{Z}_{s}^{\left(1\right)}=\sqrt{2}\left(\frac{\pi}{\sigma}\right)^{5/2}\frac{\sigma}{a} e^{-\mathcal{E}_{0}^{\left(1\right)}}.\label{eq:PF-SP1-LAWC}
\end{equation}

Hence, gathering the results in Eqs. (\ref{eq:PF-Min1-LAWC}) and
(\ref{eq:PF-SP1-LAWC}) into the first equation of (\ref{eq:LangerRR}),
we obtain

\begin{eqnarray*}
\Gamma_{\left(\pi,\pi\right)\rightarrow\left(0,\pi\right)} & = & \left|\kappa^{\left(1\right)}\right|\sqrt{\frac{\sigma}{2\pi}}\frac{\left(2+3\xi\right)\left(2+\xi\right)}{2a} e^{-\Delta\mathcal{E}^{\left(1\right)}}.
\end{eqnarray*}

At the second saddle point (\ref{eq:SP2-LAWC}) the energy is the
same, i.e., 

\begin{eqnarray*}
\mathcal{E}_{0}^{\left(2\right)} & = & \sigma\left(\frac{3}{4}\xi^{2}-1\right)
\end{eqnarray*}
whereas the energy barrier separating the minimum $\left(0,\pi\right)$
from the saddle point $(\theta_{1}^{\left(2\right)},\theta_{2}^{\left(2\right)})$ reads 

\begin{eqnarray}
\Delta\mathcal{E}^{\left(2\right)} & = & \sigma\left(1-2\xi+\frac{3}{4}\xi^{2}\right).\label{eq:EnBar2-LAWC}
\end{eqnarray}

Here the correction to the energy barrier of the free MD starts with
a negative coefficient which implies that this energy barrier decreases
when $\xi$ increases, see Fig \ref{fig:Energy-barrier}. Indeed, as the DDI becomes stronger, the first moment, which has already switched, exerts a stronger force on the
first moment. On the opposite, the second energy barrier (\ref{eq:EnBar1-LAWC})
increases with $\xi$. When the system is in the initial ferromagnetic
state $\left(\pi,\pi\right)$, it is much more difficult for the first
magnetic moment to break free from the ferromagnetic coupling.

The corresponding switching rate then reads

\begin{eqnarray*}
\Gamma_{\left(0,\pi\right)\rightarrow\left(0,0\right)} & = & \left|\kappa^{\left(2\right)}\right|\sqrt{\frac{\sigma}{2\pi}}\frac{\left(2-3\xi\right)\left(2-\xi\right)}{2a} e^{-\Delta\mathcal{E}^{\left(2\right)}}.
\end{eqnarray*}

Note that we have the symmetry $\Gamma_{\left(0,\pi\right)\rightarrow\left(0,0\right)}=\Gamma_{\left(\pi,\pi\right)\rightarrow\left(0,\pi\right)}\left(\xi\longrightarrow-\xi\right)$.
In both switching rates the prefactor $\kappa$ has been computed
numerically. The total switching rate of the MD is obtained upon
using Eq. (\ref{eq:Two-StepProc-RR}) and is plotted in Fig. \ref{fig:Gamma-DDI-LA_WCSC}.

\subsubsection{Strong coupling ($\xi\geq2/3$)}

The situation now is sketched in Fig. \ref{fig:EscapeRoute-LAWCSC} (right) with the minima

\begin{equation}
\left(\theta_{1},\theta_{2}\right)=\left(0,0\right),\left(\pi,\pi\right),\label{eq:Min-LASC}\end{equation}
 maxima

\begin{equation}
\left(\theta_{1},\theta_{2}\right)=\left(\pm\frac{\pi}{2},\pm\frac{\pi}{2}\right),\label{eq:Max-LASC}\end{equation}
 and saddle points

\begin{equation}
\left(\theta_{1},\theta_{2}\right)=\left(\pm\frac{\pi}{2},\mp\frac{\pi}{2}\right).\label{eq:SP-LASC}\end{equation}

Comparing with the WC regime, we note that the global minima remain
the same, while the metastable minima and the saddle points are no
longer stationary states. On the other hand, the anti-ferromagnetic
states given by $\left(\pm\frac{\pi}{2},\mp\frac{\pi}{2}\right)$
change as well, being maxima for the WC regime they turn into saddle
points in the SC regime, creating a switching path in which both magnetic moments switch coherently in a one-step process, as shown in Fig. \ref{fig:EscapeRoute-LAWCSC}

Following the same procedure as for WC, we obtain the expression for
the switching rate in this SC regime of DDI-MD

\begin{eqnarray}
\Gamma_{\mathrm{LASC}} & = & \alpha\tilde{\kappa}\sqrt{\frac{\sigma}{\pi}}\left(3\xi+2\right)\sqrt{\frac{2+\xi}{\xi\left(3\xi-2\right)}}e^{-\sigma\left(2+\xi\right)}\label{eq:RR-LASC}
\end{eqnarray}

where 

\begin{equation}
\tilde{\kappa}=\left(1-\frac{\xi}{2}\right)+\sqrt{\left(1+\frac{3}{2}\xi\right)^{2}+\frac{2}{\alpha^{2}}\xi\left(2+\xi\right)}.
\end{equation}

\begin{figure}
\includegraphics[width=8cm]{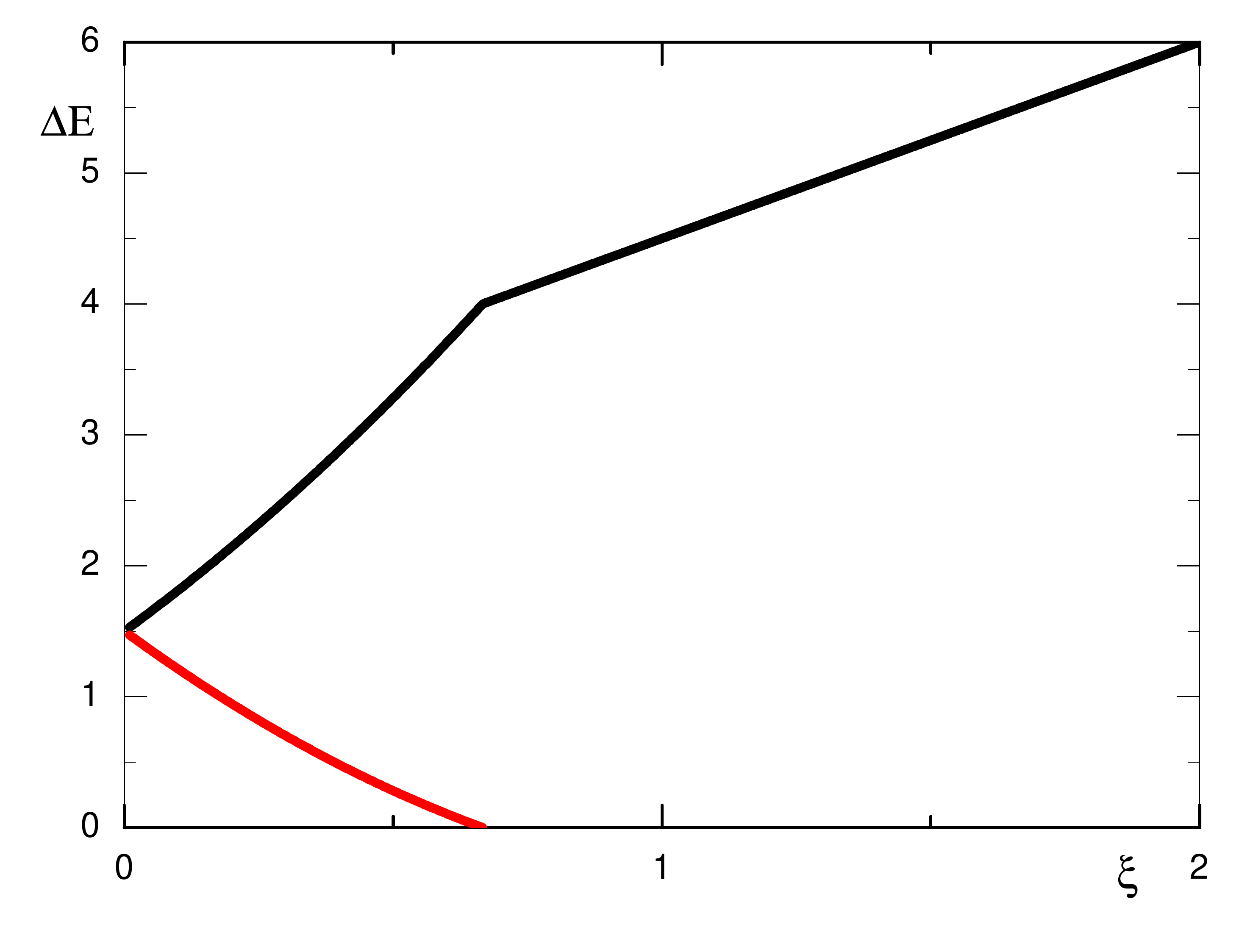}
\caption{\label{fig:Energy-barrier}Energy barrier as function of $\xi$ for the LA-MD, with $\sigma=3/2$.}
\end{figure}

Fig. \ref{fig:Energy-barrier} shows the evolution of the energy barrier
as a function of the DDI coupling $\xi$. We clearly see the two coupling
regimes separated by the critical value $\xi=2/3$. In the WC
regime we do see the two different energy barriers corresponding
to the two steps of the reversal process. The energy barrier is continuous
for all $\xi$, including the critical region. Moreover, the second step disappears at $\xi=2/3$ and the dynamics of the system
becomes a one-step process.

\begin{figure}
\includegraphics[width=8cm]{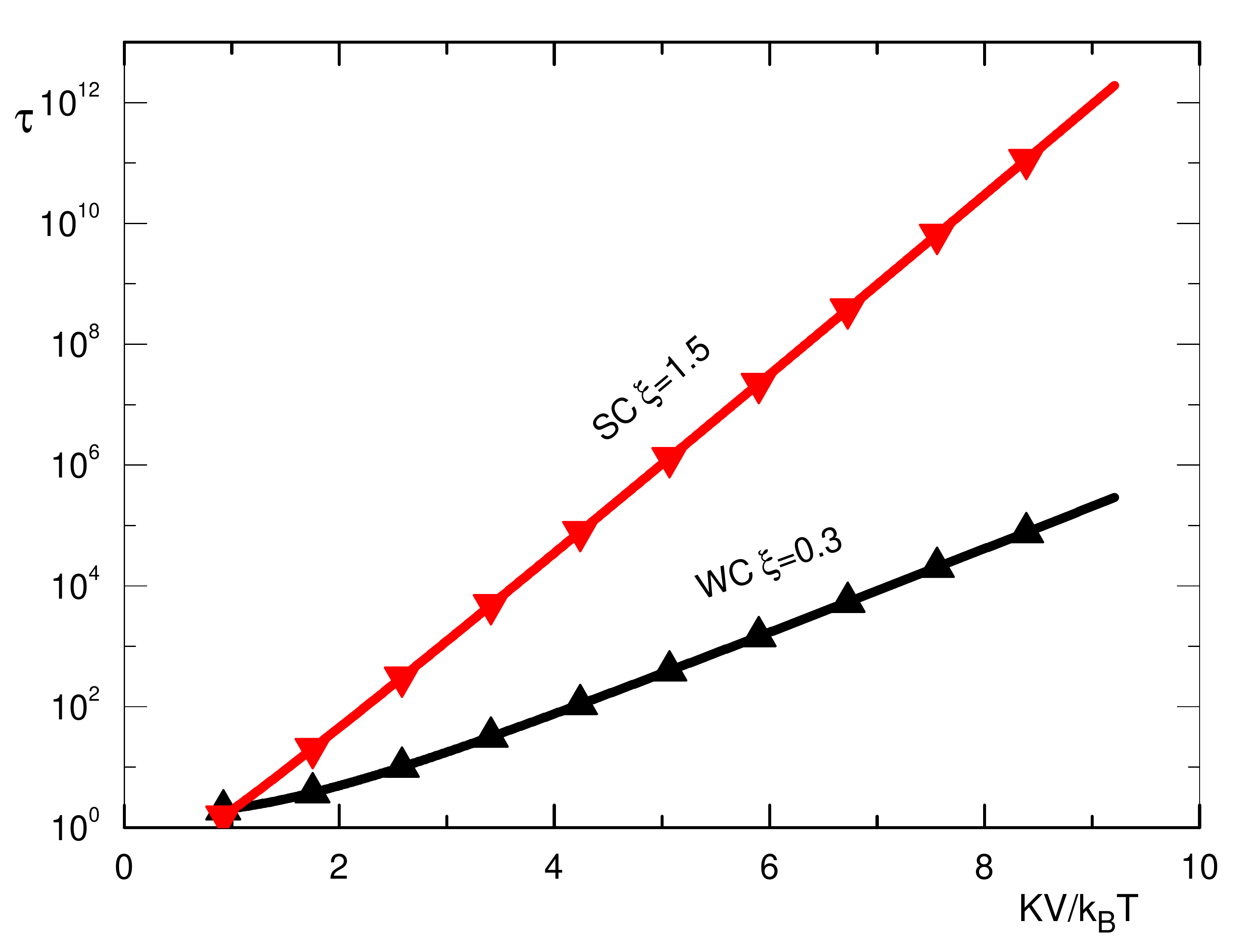}
\caption{\label{fig:Gamma-DDI-LA_WCSC}Reduced switching time $\tau$ of the DDI-MD with LA as a function of $\sigma$ for weak and strong coupling regimes .}
\end{figure}

Figure \ref{fig:Gamma-DDI-LA_WCSC} shows the behavior of the reduced switching time $\tau=1/\Gamma=t/2 t_s$ as a function of the reduced anisotropy barrier $\sigma$ for both
coupling regimes. 

The total switching time may be written as $\tau_\mathrm{WC}= \tau_{\left(\pi,\pi\right)\rightarrow\left(0,\pi\right)} + \tau_{\left(0,\pi\right)\rightarrow\left(0,0\right)}$ and it is clear that $\tau_{\left(\pi,\pi\right)\rightarrow\left(0,\pi\right)} \gg \tau_{\left(0,\pi\right)\rightarrow\left(0,0\right)}$ since in the first step one of the magnetic moments has to win against the effective FM coupling. 
Consequently, as the DDI coupling increases the magnetic moment 
switching during the first step becomes more and more difficult to achieve and thereby the corresponding switching time increases, which explains why 
$\tau_\mathrm{LAWC} < \tau_\mathrm{LASC}$.

\subsection{Transverse anisotropy}

\begin{figure*}
\includegraphics[width=16cm]{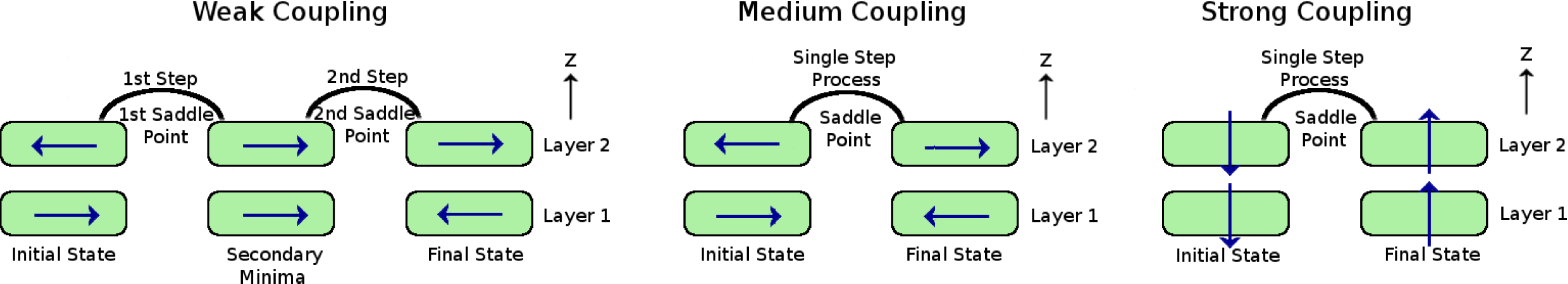}
\caption{Escape route for the three DDI coupling regimes with TA configuration\label{fig:EscapeRoute-TAWCMCSC}}
\end{figure*}

In spherical coordinates the energy now reads

\begin{eqnarray}
\mathcal{E} & = & -\sigma\left(\sin^{2}\theta_{1}+\sin^{2}\theta_{2}\right) \\
& & -\sigma\xi\left[\cos\left(\theta_{1}+\theta_{2}\right)+\cos\theta_{1}\cos\theta_{2}\right]\nonumber\label{eq:En-TA-SC}
\end{eqnarray}

In this situation, we find that there are two minima with the corresponding
lowest eigenvalues $\sigma\left(2-\xi\right)$ and $\sigma\left(2-3\xi\right)$
whose vanishing leads to the two critical DDI couplings $\xi=2$ and $\xi=2/3$. 

\subsubsection{Weak coupling$\left(\xi\leq2/3\right)$}

The absolute minima of the system are now

\begin{equation}
\left(\theta_{1},\theta_{2}\right)=\left(\pm\frac{\pi}{2},\mp\frac{\pi}{2}\right)\label{eq:Min-TAWC}
\end{equation}

while the local minima are given by

\begin{equation}
\left(\theta_{1},\theta_{2}\right)=\left(\pm\frac{\pi}{2},\pm\frac{\pi}{2}\right).\label{eq:LMin-TAWC}
\end{equation}

On the other hand, the absolute maxima are the anti-ferromagnetic
states

\begin{equation}
\left(\theta_{1},\theta_{2}\right)=\left(0,\pi\right),\left(\pi,0\right)\label{eq:Max-TAWC}
\end{equation}
while the local maxima correspond to the ferromagnetic states
\begin{equation}
\left(\theta_{1},\theta_{2}\right)=\left(0,0\right),\left(\pi,\pi\right).\label{eq:LMax-TAWC}
\end{equation}

Finally, the saddle points are

\begin{equation}
\left(\theta_{1},\theta_{2}\right)=\left(\varepsilon_{1}\arccos(\varepsilon_{2}x_{1}^{\gamma}),\varepsilon_{1}\arccos(\varepsilon_{2}x_{2}^{\gamma})\right).\label{eq:SP-TAWC}
\end{equation}

The escape route is sketched in Fig. \ref{fig:EscapeRoute-TAWCMCSC} (left).

\begin{figure*}
\subfloat[Weak Coupling $\xi=0.2$\label{fig:Energyscapes-weak-transversal}]{\includegraphics[width=6cm]{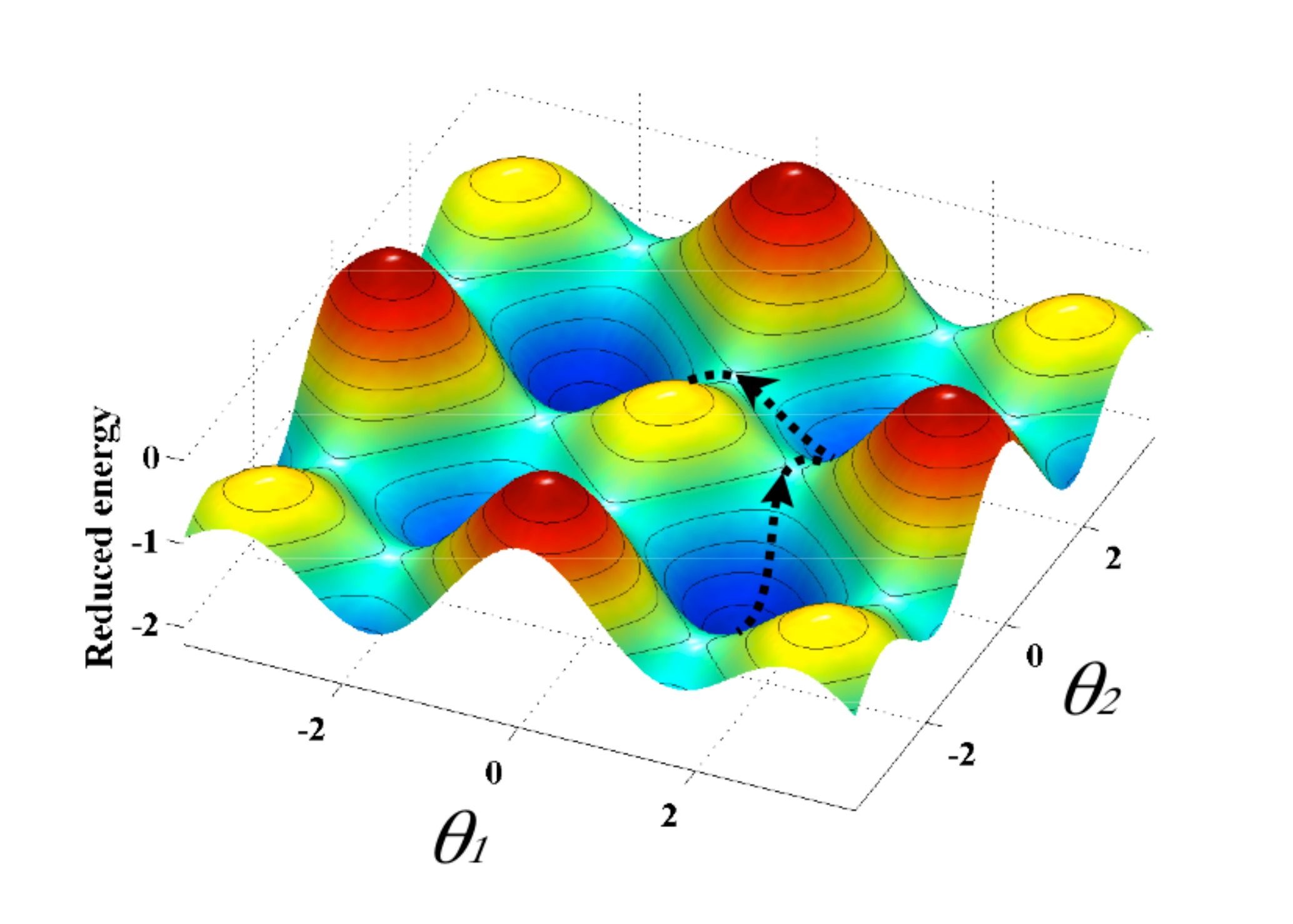}}
\subfloat[Medium coupling $\xi=1$\label{fig:Energyscapes-medium-transversal}]{\includegraphics[width=6cm]{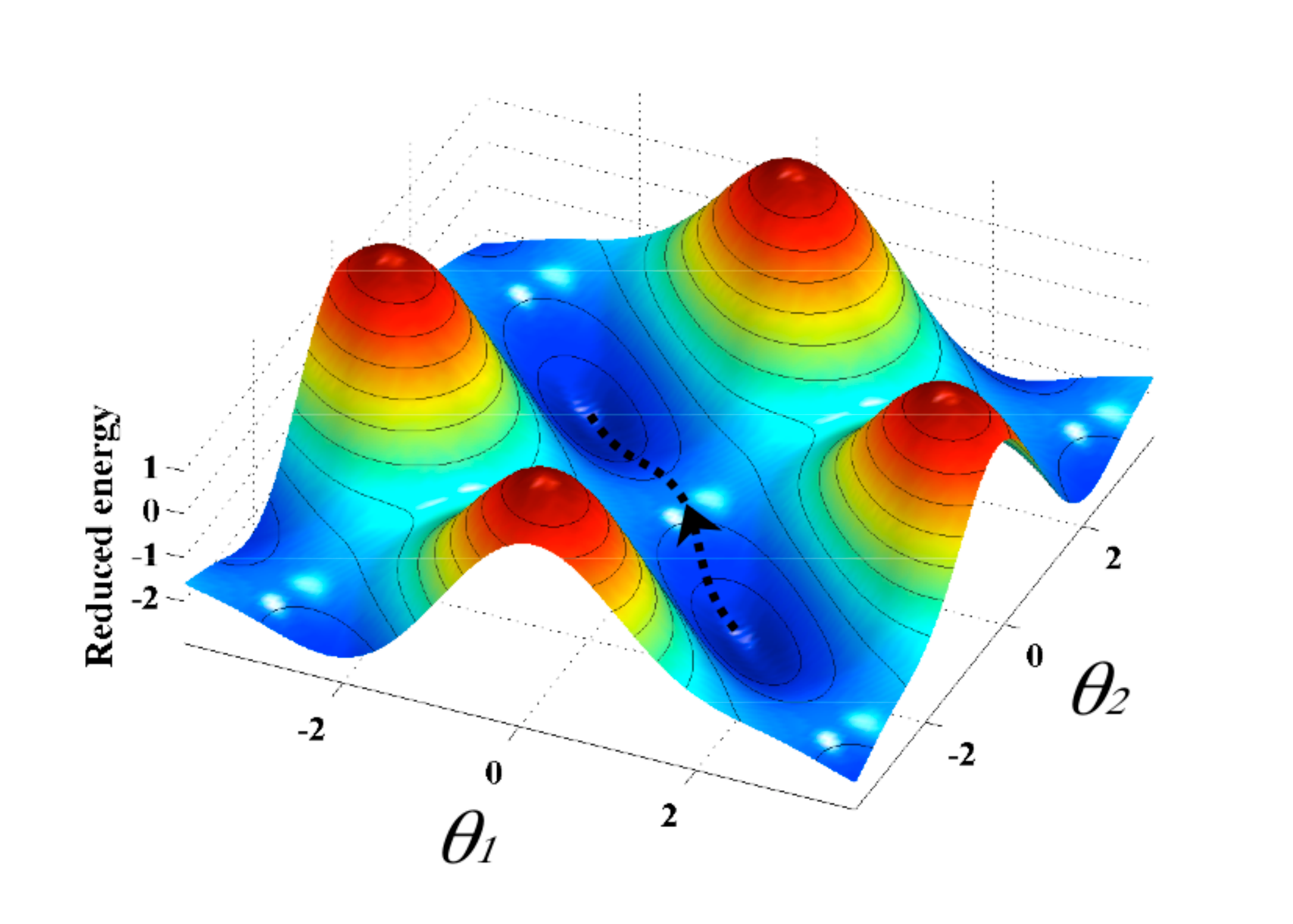}}
\subfloat[Strong Coupling $\xi=6.33$\label{fig:Energyscapes-strong-transversal}]{\includegraphics[width=6cm]{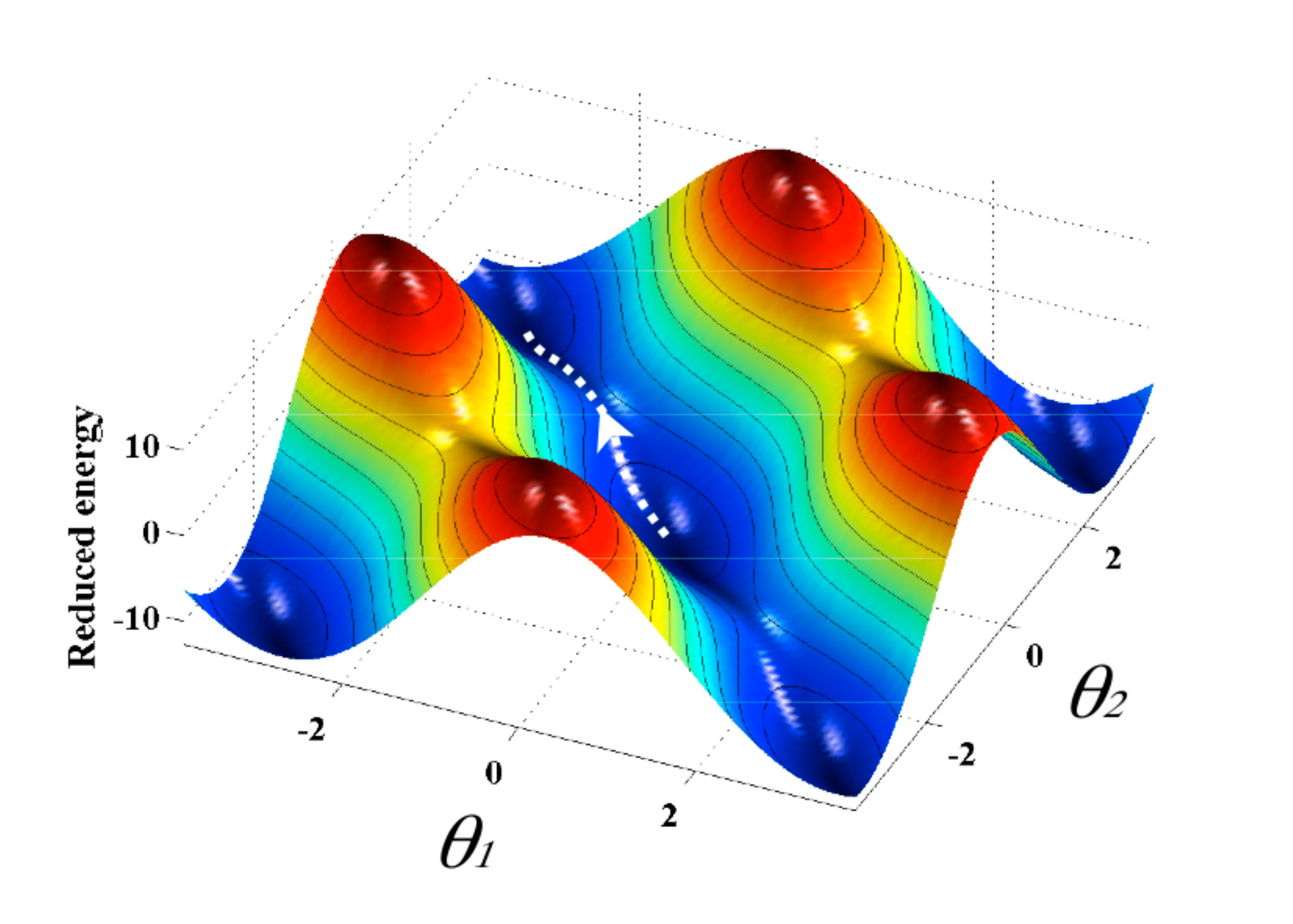}}
\caption{\label{fig:Energyscapes-transversal}Energy potential surface for the different regimes of the MD with TA and identical $\sigma=3/2$.}
\end{figure*}

In the present case, as can be seen in Fig. \ref{fig:Energyscapes-transversal},
the initial state is given by $(\frac{\pi}{2},-\frac{\pi}{2})$, the
metastable minimum is $(\frac{\pi}{2},\frac{\pi}{2})$, and
the first and second saddle points are given by $(\arccos x_{1}^{+},\arccos x_{2}^{+})$
and $(\arccos x_{1}^{-},\arccos x_{2}^{-})$, respectively. 
In this weak coupling regime, the MD switching is again a two-step
process and the corresponding switching rates are given by

\begin{widetext}
\begin{eqnarray}
\Gamma_{\mathrm{TAWC}}^{(1)} & = & \left|\frac{\kappa^{(1)}}{\xi}\right|\frac{1}{2\sqrt{3}\pi a}\sqrt{\left(3\xi+2\right)
\left(1+\xi\right) \left(2-\xi \right)}\times e^{-\frac{3\sigma}{2}\left(\frac{2}{3}+\xi\right)\left(1-\frac{\xi}{2}\right)}\nonumber\\
\Gamma_{\mathrm{TAWC}}^{(2)} & = & \Gamma_{\mathrm{TAWC}}^{(1)} \left( \xi \rightarrow -\xi \right) \label{eq:RR-TAWC}
\end{eqnarray}
\end{widetext}
where the attempt frequencies $\kappa^{(i)},i=1,2$ are calculated numerically; $a$ is defined in Eq. (\ref{eq:Roots-Const_a}).
The total switching rate is given by Eq. (\ref{eq:Two-StepProc-RR}).

\subsubsection{Medium coupling ($\frac{2}{3}<\xi<2$)}

The minima are given by
\begin{equation}
\left(\theta_{1},\theta_{2}\right)=\left(\pm\frac{\pi}{2},\mp\frac{\pi}{2}\right),\label{eq:Min-TAMC}
\end{equation}
the maxima correspond to the anti-ferromagnetic states

\begin{equation}
\left(\theta_{1},\theta_{2}\right)=\left(0,\pi\right),\left(\pi,0\right)\label{eq:Max-TAMC},
\end{equation}
and the saddle points are the ferromagnetic states

\begin{equation}
\left(\theta_{1},\theta_{2}\right)=\left(0,0\right),\left(\pi,\pi\right)\label{eq:SP1-TAMC}
\end{equation}
and

\begin{equation}
\left(\theta_{1},\theta_{2}\right)=\left(\pm\frac{\pi}{2},\pm\frac{\pi}{2}\right).\label{eq:SP2-TAMC}
\end{equation}

Starting from the initial state $(\frac{\pi}{2},-\frac{\pi}{2})$,
the switching is a single-step process in which the two magnetic moments
switch coherently through the saddle point $(0,0)$ leading
into the state $(-\frac{\pi}{2},\frac{\pi}{2})$, as shown in Fig.
\ref{fig:EscapeRoute-TAWCMCSC} (middle).

The switching rate for this coupling regime reads
\begin{equation}
\Gamma_{\mathrm{TAMC}}=\frac{\alpha\left(\xi-2\right)}{\pi}\sqrt{\frac{\left(3\xi+2\right)\left(1+\xi\right)}{3\xi^{2}\left(3\xi-2\right)}}e^{-\sigma\left(2-\xi\right)}\label{eq:RR-TAMC}
\end{equation}
where the attempt frequency has been obtained analytically and is
given by $\kappa=2\alpha\left(\xi-2\right)$.

\subsubsection{Strong coupling ($\xi>2$)}
Here the DDI field is twice larger than the anisotropy field and the minima are the ferromagnetic states
\begin{equation}
\left(\theta_{1},\theta_{2}\right)=\left(0,0\right),\left(\pi,\pi\right),\label{eq:Min-TASC}
\end{equation}
the maxima are the anti-ferromagnetic states
\begin{equation}
\left(\theta_{1},\theta_{2}\right)=\left(0,\pi\right),\left(\pi,0\right)\label{eq:Max-TASC},
\end{equation}
and the saddle points are located at
\begin{equation}
\left(\theta_{1},\theta_{2}\right)=\left(\pm\frac{\pi}{2},\pm\frac{\pi}{2}\right),\left(\pm\frac{\pi}{2},\mp\frac{\pi}{2}\right).\label{eq:SP-TASC}
\end{equation}

As the DDI coupling increases and the system enters the strong coupling
regime, the states that previously were minima in the medium coupling
regime become saddle points, and \emph{vice versa}. The system still
relaxes in a one step process, but this time from the initial state $(0,0)$ through the saddle point $(-\frac{\pi}{2},\frac{\pi}{2})$. This is sketched in Fig.\ref{fig:EscapeRoute-TAWCMCSC} (right).
The switching rate in this case reads 
\begin{equation}
\Gamma_{\mathrm{TASC}}=\frac{\left|\kappa\right|}{\pi}\sqrt{\frac{3\xi^{2}\left(3\xi-2\right)}{\left(2+3\xi\right)
\left(1+\xi\right)}} e^{-\sigma\left(\xi-2\right)}\label{eq:RR-TASC}
\end{equation}

with 

\begin{equation*}
\kappa=\alpha\left[\left(4+\xi\right)-\sqrt{9\xi^{2}+\frac{8}{\alpha^{2}}\left(\xi^{2}-\xi-2\right)}\right]. 
\end{equation*}

\begin{figure}
\includegraphics[width=8cm]{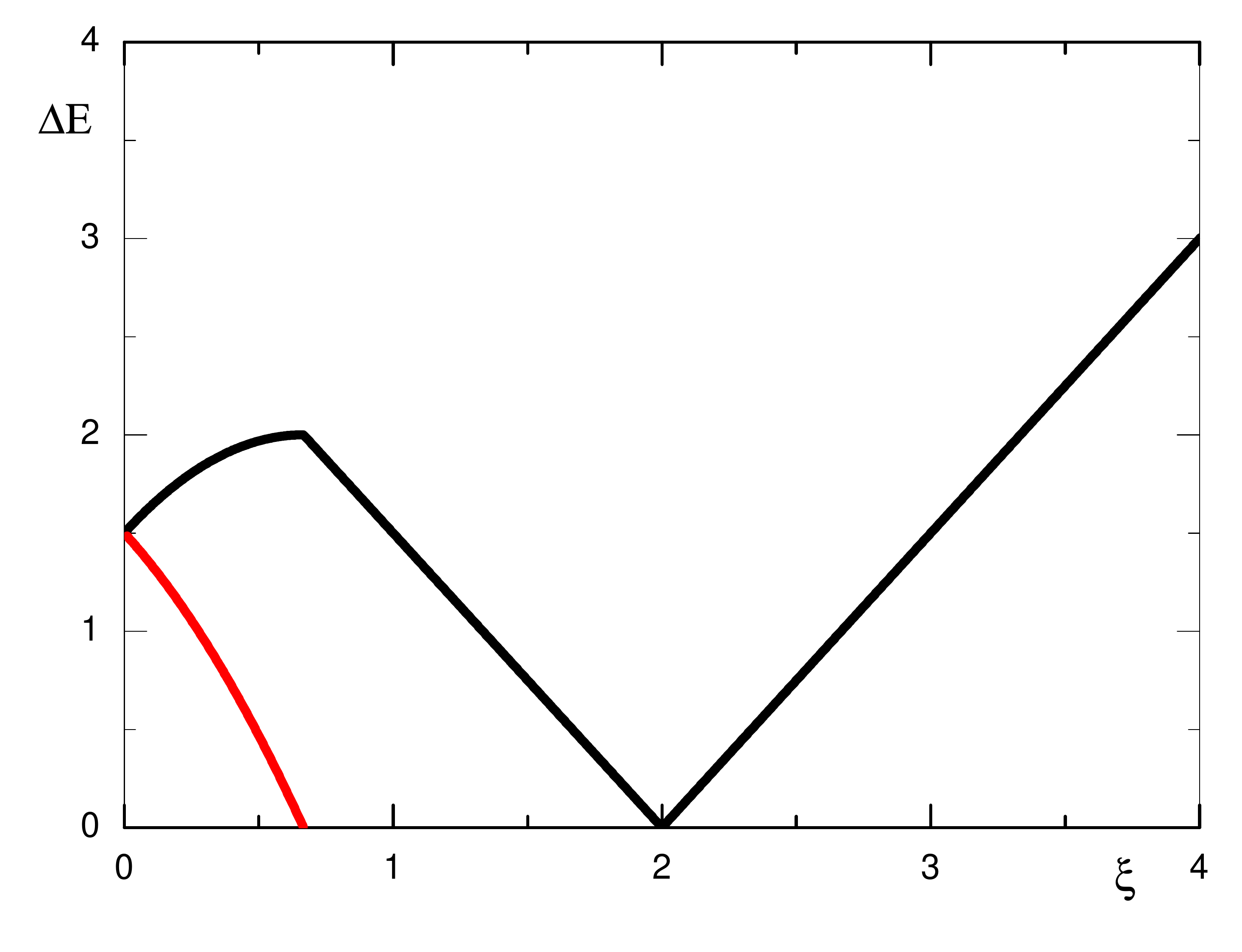}
\caption{\label{fig:energy-barrier}Energy barrier of the MD with TA as a function of $\xi$, $\sigma=3$.}
\end{figure}

\begin{figure}
\includegraphics[width=8cm]{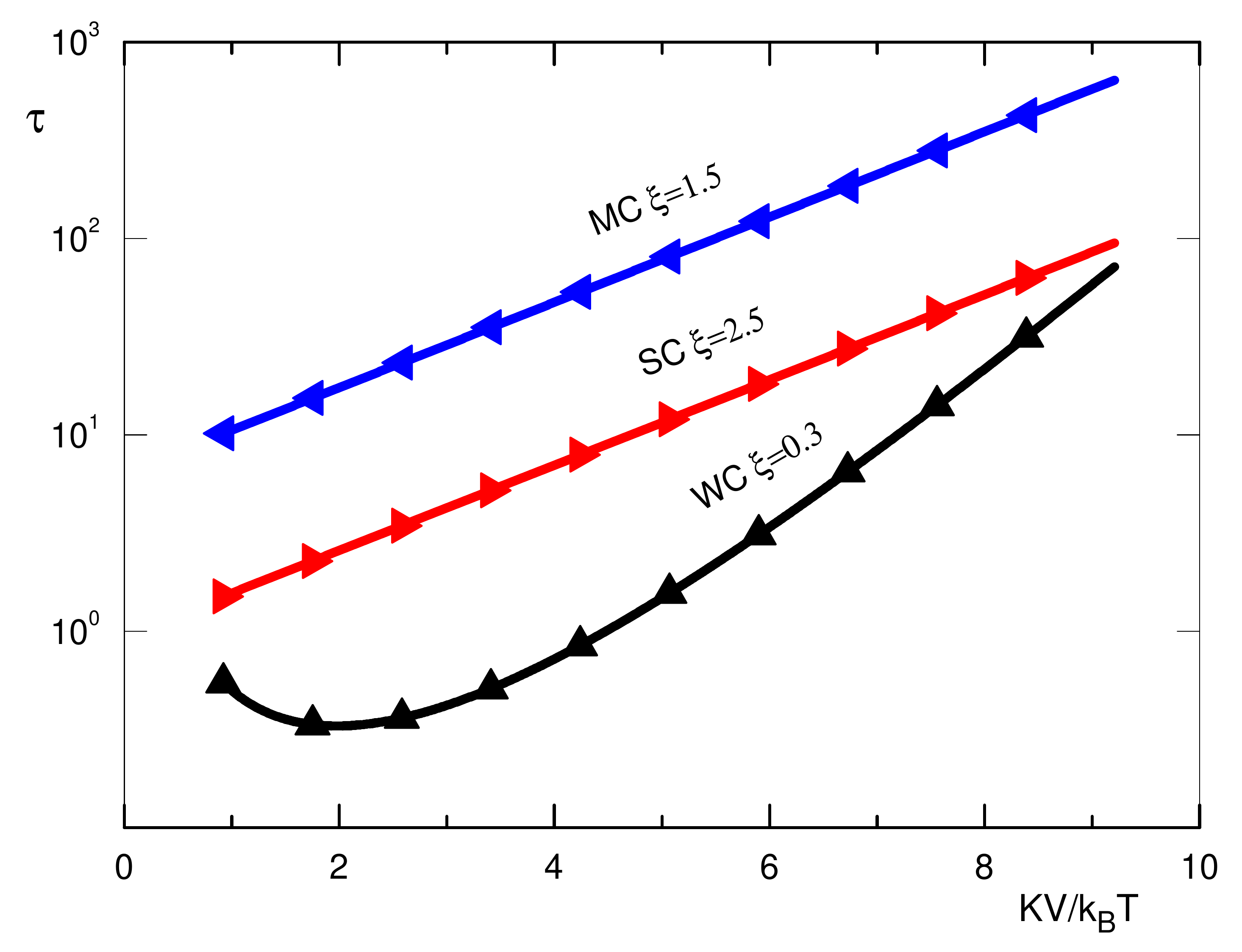}
\caption{\label{fig:relax-rate}Reduced switching time $\tau$ of the MD with TA for the three different coupling regimes as a function of $\sigma$.}
\end{figure}

Fig. \ref{fig:energy-barrier} shows the evolution of the energy barrier
as $\xi$ changes. Similarly to Fig. \ref{fig:Energy-barrier}, the
different regimes are clearly identified, so are the continuity of
the barrier and the disappearance of the second step at $\xi=2/3$.
At $\xi=2$, we can see that the energy barrier vanishes forming a
{}``furrow'' that connects directly (with no energy barriers) the
states $(0,0)$, $(\pi,\pi)$, and $(\pm\frac{\pi}{2},\mp\frac{\pi}{2})$. 
Figure \ref{fig:relax-rate} shows the behavior of the switching
time in the three regimes of DDI coupling, namely WC, MC and SC; it
is a plot of the inverse of the expressions (\ref{eq:RR-TAWC}), (\ref{eq:RR-TAMC}),
and (\ref{eq:RR-TASC}), respectively. 

We again observe that the SC  switching time is longer than that of the WC, similar to
what we have observed in the case of LA, see Fig.\ref{fig:Gamma-DDI-LA_WCSC}.
Furthermore, we see that, as a function of $\sigma\sim 1/T$, the (logarithm of)
the switching time in the WC regime is not a straight line.
This implies that the prefactor plays a dominant role. As $\xi$ increases the switching time
becomes dominated by the Arrhenius (exponential) law where the prefactor is a constant thus
leading to a straight line in a logarithmic plot of the switching rate as a function of $\sigma$.
One of the consequences of a dominant prefactor is that the switching rate becomes
quite sensitive to damping and thus to the coupling of the system to its thermal bath and to the
various fluctuations.

\subsection{Mixed anisotropy}

\begin{figure*}
\subfloat[Weak Coupling $\xi=0.2$\label{fig:Energyscapes-weak-mixed}]{\includegraphics[width=6cm]{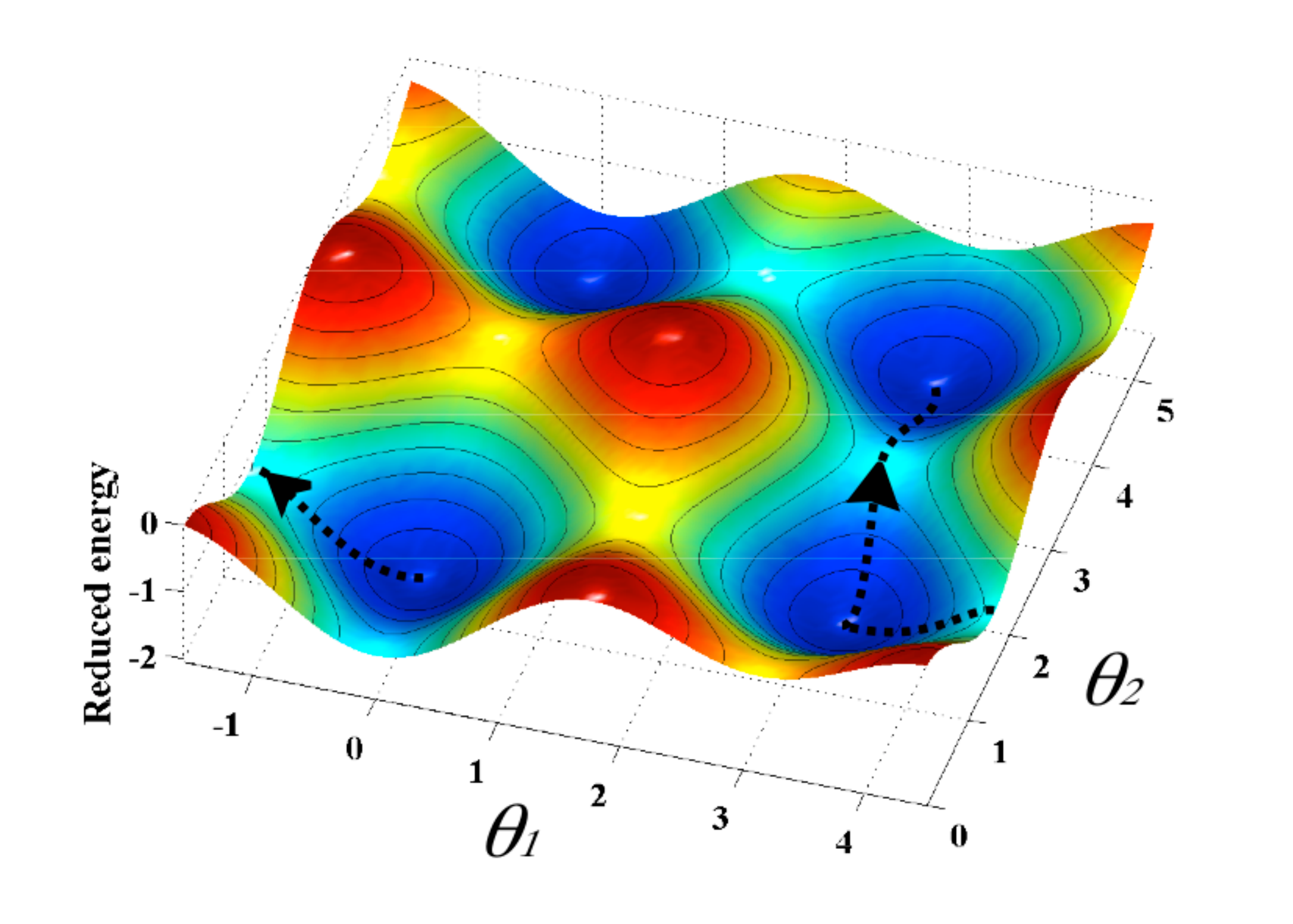}}
\subfloat[Approximate escape route WC\label{fig:relax-path-weak-mixed}]{\includegraphics[width=4cm]{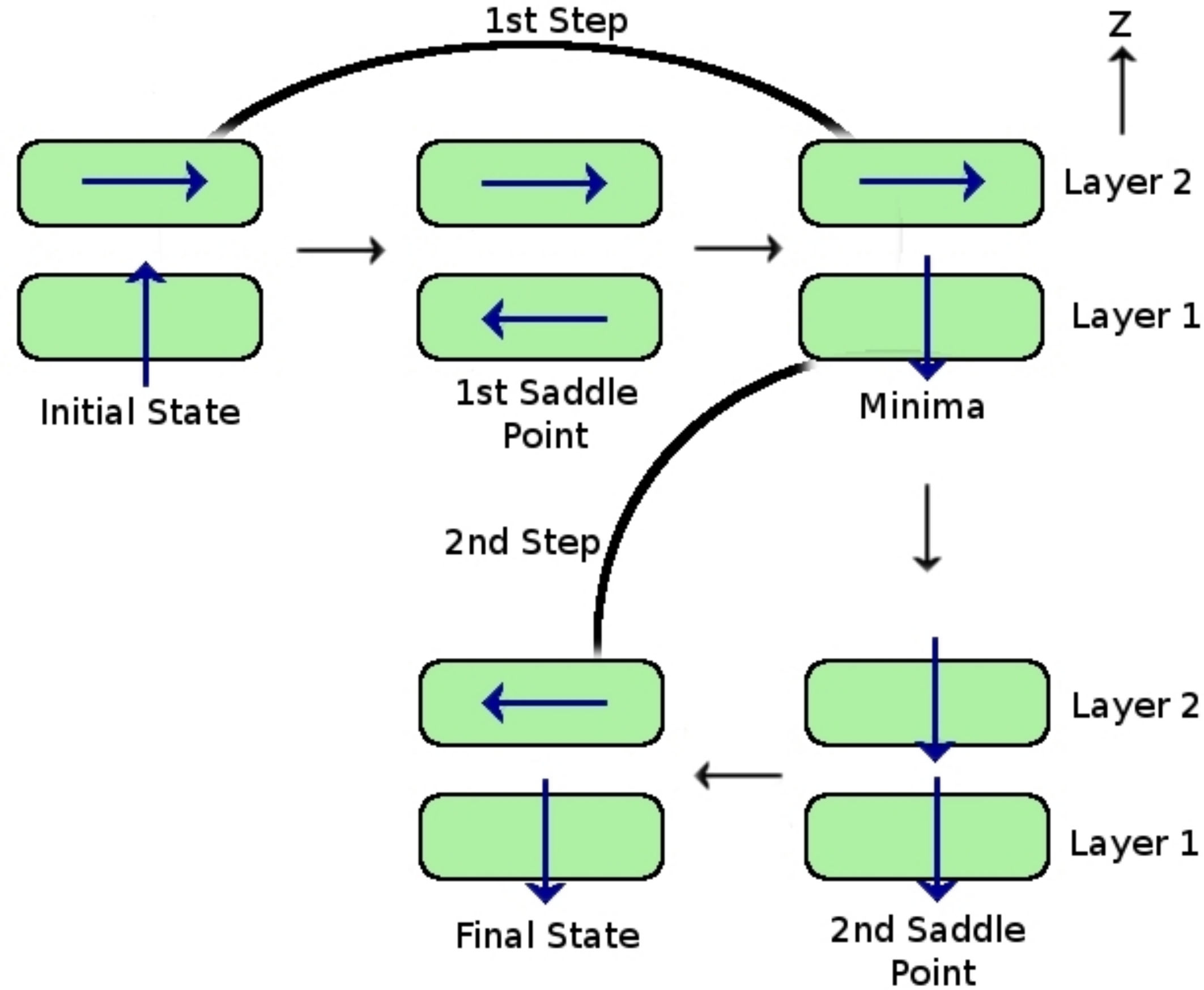}}
\subfloat[Strong Coupling $\xi=1.66$\label{fig:Energyscapes-strong-mixed}]{\includegraphics[width=6cm]{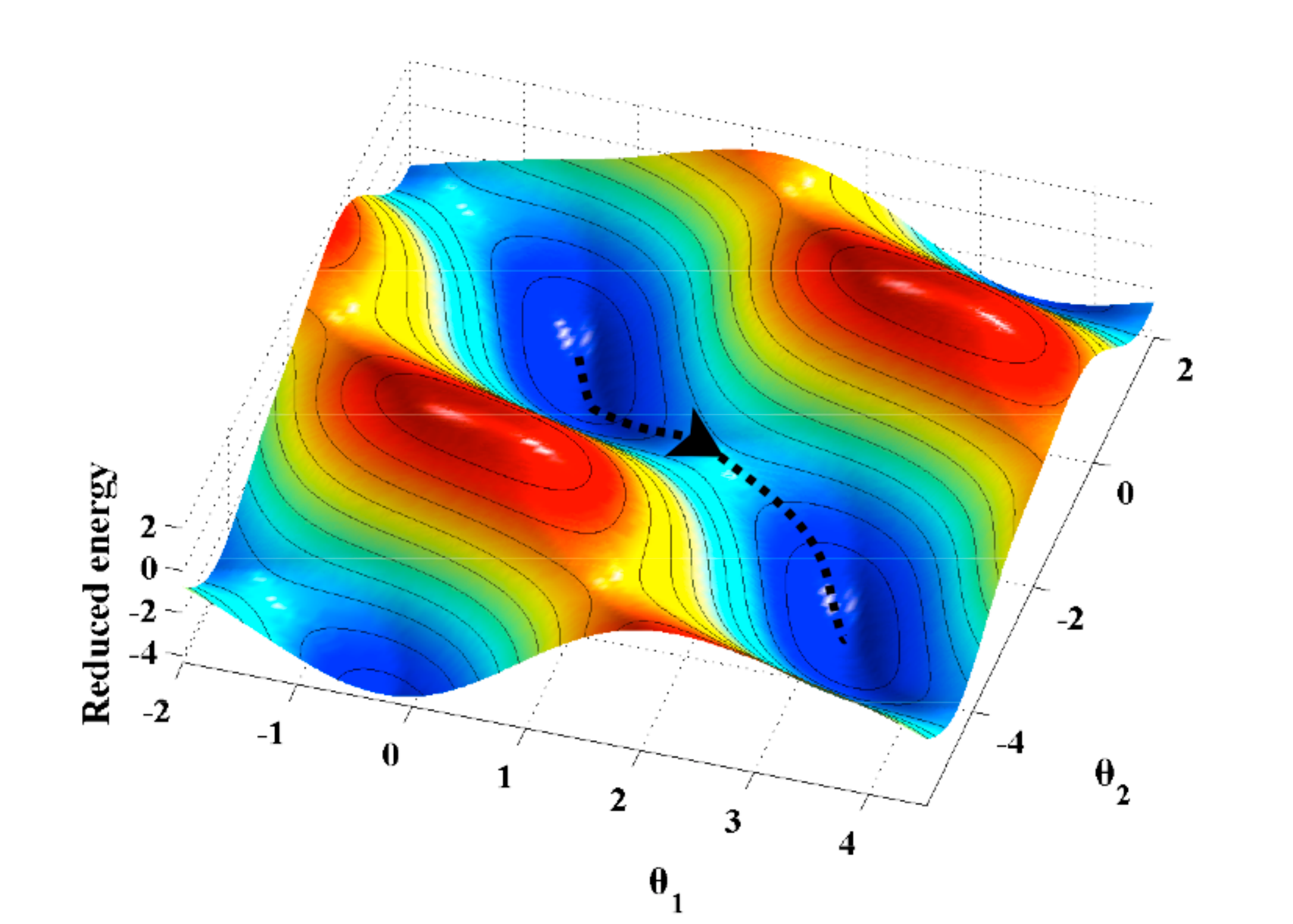}}
\caption{\label{fig:Energyscapes-mixed}Energy potential surface for the different regimes of the MD and WC escape route with MA and $\sigma=\frac{3}{2}$.}
\end{figure*}

Whereas the LA and TA configurations for DDI coupling have a somewhat
similar behavior in terms of energy potential surface evolution and coupling regimes,
the MA shows a completely different and more complex behavior, as
can be seen in Fig. \ref{fig:Energyscapes-mixed}. Analysis of the
stationary states reveals the presence of two coupling regimes as
well, but with a critical value now at $\xi=\frac{2}{\sqrt{3}}$.
As before, switching in the WC regime is again a two-step process while it is a one-step process in the SC regime. 
The stationary states are given by Eq. \ref{eq:SS-WC-DDI-MA} together with
\begin{eqnarray*}
\left(\theta_{1},\theta_{2}\right) & = & \left(0,\pm\pi\right),\left(\pm\pi,0\right),\left(\pm\pi,\pm\pi\right),\\
 &  & \left(\pm\pi,\mp\pi\right),\left(\pm\frac{\pi}{2},\pm\frac{\pi}{2}\right),\left(\pm\frac{\pi}{2},\mp\frac{\pi}{2}\right).
\end{eqnarray*}

It is worth investigating this case because it corresponds to another interesting situation with a rather thick layer coupled to a rather thin layer. It may also be relevant in the situation of a soft magnetic layer coupled to another hard magnetic layer.

The remaining stationary states are rather cumbersome and they are relegated to the appendix.

\subsubsection{Weak coupling($\xi\leq\frac{2}{\sqrt{3}}$)}

The individual switching rates are given by Eq. (\ref{eq:RR-MAWC}), 
and the full switching rate is given by Eq. (\ref{eq:Two-StepProc-RR}). The corresponding
analytical expressions are too cumbersome and are thus given in the appendix. They are plotted in Fig. \ref{fig:RR-MA}. In this case too, $\kappa$ is computed numerically.

\subsubsection{Strong coupling (\texorpdfstring{$\xi>\frac{2}{\sqrt{3}}$}{})}
The minima now are the FM states
\begin{equation*}
\left(\theta_{1},\theta_{2}\right)=\left(0,0\right),\left(\pi,\pi\right) 
\end{equation*}
and the maxima are the AFM states
\begin{equation*}
\left(\theta_{1},\theta_{2}\right)=\left(0,\pi\right),\left(\pi,0\right)
\end{equation*}
while the saddle points are
\begin{equation*}
\left(\theta_{1},\theta_{2}\right)=\left(\pm\frac{\pi}{2},\pm\frac{\pi}{2}\right),\left(\pm\frac{\pi}{2},\mp\frac{\pi}{2}\right).
\end{equation*}

Starting in the WC regime, as the DDI coupling increases, the minima given by
\begin{equation*}
\left(\pm(\arccos x_{+}^{-}-\arccos x_{-}^{-})/2,\pm(\arccos x_{+}^{-}+\arccos x_{-}^{-})/2\right)
\end{equation*}
start to merge into the saddle point $(0,0)$. When the system enters
the SC coupling regime, these two minima completely merge leading
to the transformation of the state $(0,0)$ from a saddle point into
a minimum. The minima 
\begin{eqnarray*}
\left(\pm\pi \right. & \mp & (\arccos x_{+}^{-}+\arccos x_{-}^{-})/2,\\
\pm\pi & \mp & \left.(\arccos x_{+}^{-}-\arccos x_{-}^{-})/2\right)
\end{eqnarray*}
have a similar behavior around the state $\left(\pi,\pi\right)$,
while the saddle points $\left(\pm\frac{\pi}{2},\mp\frac{\pi}{2}\right)$
do not change. The escape route in the SC regime is similar to
that of the SC for LA and TA. However, the shape of the energy potential surface
presents curved paths instead of the usual straight paths, see Fig. \ref{fig:Energyscapes-mixed}. 

The escape rate from the initial state $\left(0,0\right)$ through the saddle
point $(-\frac{\pi}{2},\frac{\pi}{2})$ is given by
\begin{eqnarray}
\Gamma_{\mathrm{MASC}} & = & \frac{\left|\kappa\right|}{\pi}e^{-\sigma\xi}\times\nonumber\\
&
&\sqrt{\frac{\left[\left(1+2\xi\right)^{2}-r^{2}\right]\left[4\xi^{2}-r_{p}^{2}\right]}{\left[
\left(1+\xi\right)^{2}-r^{2}\right]\left[\left(\xi+2r\right)\left|\xi-2r\right|\right]}},
\label{eq:RR-MASC} 
\end{eqnarray}
where $r = \sqrt{1 + \xi^2},r_p=\sqrt{4 + \xi^2}$. $\kappa$ is computed numerically.
\begin{figure}
\includegraphics[width=8cm]{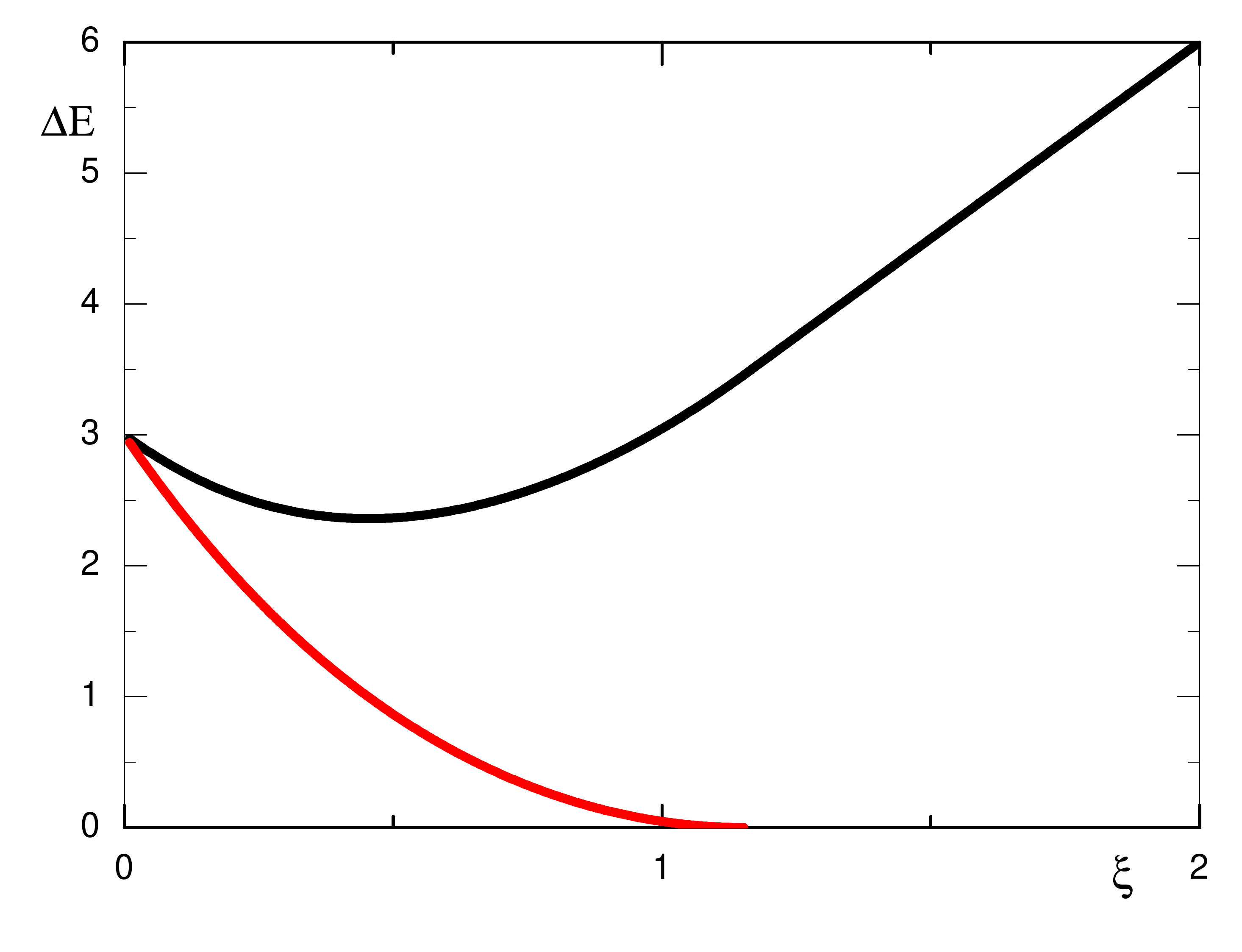}
\caption{\label{fig:energy-barrier-Axis-Plane}Changes of the energy barrier as a function of $\xi$ with MA and $\sigma=3$.}
\end{figure}

\begin{figure}
\includegraphics[width=8cm]{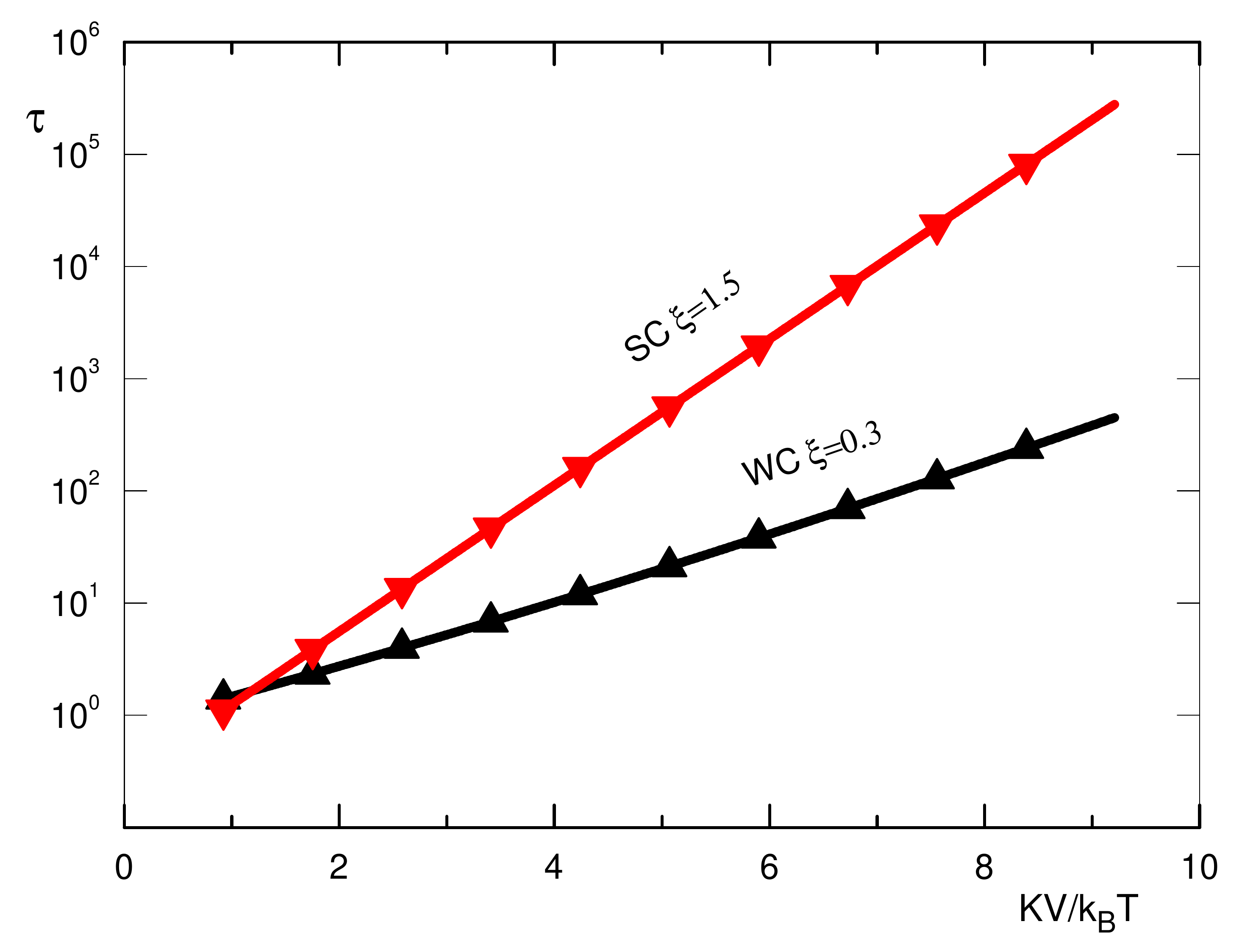}
\caption{\label{fig:RR-MA}Reduced switching time $\tau$ of the DDI-MD with MA for weak and strong
coupling regimes.}
\end{figure}

Fig. \ref{fig:energy-barrier-Axis-Plane} shows the energy barrier as a function of $\xi$.
The energy barrier along the second step of the process disappears at the critical value
$\xi=2/\sqrt{3}$, where the dynamics of the system becomes a one-step process.
As the DDI increases to higher values, the energy barrier increases,
leading to a constant increase in the switching time for high $\xi$,
as can be seen in Fig. \ref{fig:RR-MA}, where the switching time
for SC is much higher than that of WC.

By way of comparison we gather in Table \ref{tab:DDI} some results for the DDI-MD.
\begin{table}
\begin{tabular}{|c|c|c|c|}
\hline 
$t/2 t_s$($\sigma=5$) & LA & TA & MA\tabularnewline 
\hline 
\hline 
WC & $355.83$ & $1.48$ & $20.29$\tabularnewline
\hline 
SC & $1.04\times10^{6}$ & $11.57$ & $505.99$\tabularnewline
\hline
\end{tabular}
\caption{\label{tab:DDI} Reduced switching times for the WC and SC for the three anisotropy
configurations. $\xi=0.3$ for WC. $\xi=1.5$ for LA-SC and MA-SC. $\xi=2.5$ for TA-SC}
\end{table}
From this table, for $\sigma=5$, we see that for the three anisotropy configurations the
switching time increases then decreases with the DDI coupling $\xi$. The fastest
switching seems to occur for weak coupling and transverse anisotropy. 
This means that a magnetic dimer composed of two rather thin films should exhibit the fastest dynamics.

\begin{table}
\begin{tabular}{|c|c|}
 \hline
  $t$($\sigma=8$, $\xi=0.346 $) 
  & \begin{tabular}{c|c|c}
    LA & TA & MA\tabularnewline 
    \hline 
    \hline 
    $3.42\times10^{-6} s$ & $9.84\times10^{-10} s$ & $7.07\times10^{-9} s$\tabularnewline
  \end{tabular} \tabularnewline
  \hline
\end{tabular}
\caption{\label{tab:DDIAtomic} Switching times for the atomic Cobalt dimer for each of the three anisotropy
configurations.}
\end{table}

Table \ref{tab:DDIAtomic} shows the switching times for cobalt atoms in the three different anisotropies. The parameters used for the calculations are $ \mu_a=1.57\times 10^{-23} A m^{-1} atom^{-1}$, $K_a = 2.53\times 10^{-24} J atom^{-1}$, 
 $d=2\times 1.52\times 10^{-10} m$ and  $t_s=1.76\times10^{-11} s$.

It is clearly seen that in this case TA leads to the shortest switching time.
As such, in a chain of atoms a magnetic excitation should propagate faster if the anisotropy is normal to the chain axis.

\section{DM coupled magnetic dimer}

As discussed in section \ref{sub:Energy}, DMI is also relevant in
the present study and is investigated on the same footing as EI and DDI. 
Its effect is compared to that of the latter on the switching mechanisms
of the MD. In order to investigate the effect of pure DMI, we consider
the energy in Eq. (\ref{eq:DM-Energy}) without the magnetic field
and without the EI and DDI. In Ref. \citep{crelac98jmmm} it was shown
that for a simple cubic lattice, on the $(1\,0\,0)$ surface the DMI
vector $\mathbf{D}$ lies in the layer plane and thus induces perpendicular
anisotropy. 
Accordingly, in Eq. (\ref{eq:DM-Energy}) we drop the Zeeman energy,
the EI and DDI contributions. We consider two situations where the DMI vector $\mathbf{D}$
lies in the MD plane and the anisotropy easy axes parallel or perpendicular to it.

\subsection{$\mathbf{D}$ parallel to the anisotropy axes}

After simplification, the reduced energy is a function of only the
polar angles $\theta_{1},\theta_{2}$ and is given by
\begin{equation}
\varepsilon(\theta_{1},\theta_{2})=-k\left(\cos^{2}\theta_{1}+\cos^{2}\theta_{2}\right)+\delta\sin\theta_{1}\sin\theta_{2}.\label{eq:MDEnergy-DMI}
\end{equation}

The anisotropy parameter $k=0,1$ is simply a switch introduced so as to be able to keep track of the
anisotropy contribution in the subsequent results. 
Analysis of the stationary points yields
\begin{eqnarray*}
\varepsilon(0,0)=\varepsilon(\pi,\pi) & = & \varepsilon(0,\pi)=\varepsilon(\pi,0)=-2k,\\
\varepsilon(\pm\frac{\pi}{2},\pm\frac{\pi}{2}) & = & \delta,\quad \varepsilon(\pm\frac{\pi}{2},\mp\frac{\pi}{2}) = -\delta,\\
\varepsilon(\pm\frac{\pi}{2},\mp\arcsin\left(\frac{\delta}{2k}\right)) & = &
-k\left[1 + \left(\frac{\delta}{2k}\right)^{2}\right].
\end{eqnarray*}

From this analysis, we find that there is a critical value for the
DMI which separates the weak and the strong coupling regimes. In our
normalization with respect to the anisotropy energy {[}see Eq. (\ref{eq:DimParamsMat}){]}
this critical value is $\delta/k=2$, see Fig. \ref{fig:RT-En-DMI-MD-LA} (left). 
\subsubsection{Weak coupling $\delta/k<2$}
The minima are the FM and AFM states in the direction of anisotropy
\begin{equation*}
\left(\theta_{1},\theta_{2}\right)=\left(0,0\right),\left(\pi,\pi\right),\left(0,\pi\right),
\left(\pi,0\right),
\end{equation*}
the maxima are at 
\begin{equation*}
\left(\theta_{1},\theta_{2}\right)=\left(\pm\frac{\pi}{2},\pm\frac{\pi}{2}\right),
\end{equation*}
and the saddle points are
\begin{eqnarray*}
\left(\theta_{1}^{(s)},\theta_{2}^{(s)}\right) & = & \left(\pm\frac{\pi}{2},\mp\arcsin\left(\frac{\delta}{2k}\right)\right),\\
\left(\theta_{1}^{(s)},\theta_{2}^{(s)}\right) & = & \left(\mp\arcsin\left(\frac{\delta}{2k}\right),\pm\frac{\pi}{2}\right).
\end{eqnarray*}

\begin{figure*}
\subfloat[Weak Coupling $\delta=1$]{\includegraphics[width=7cm]{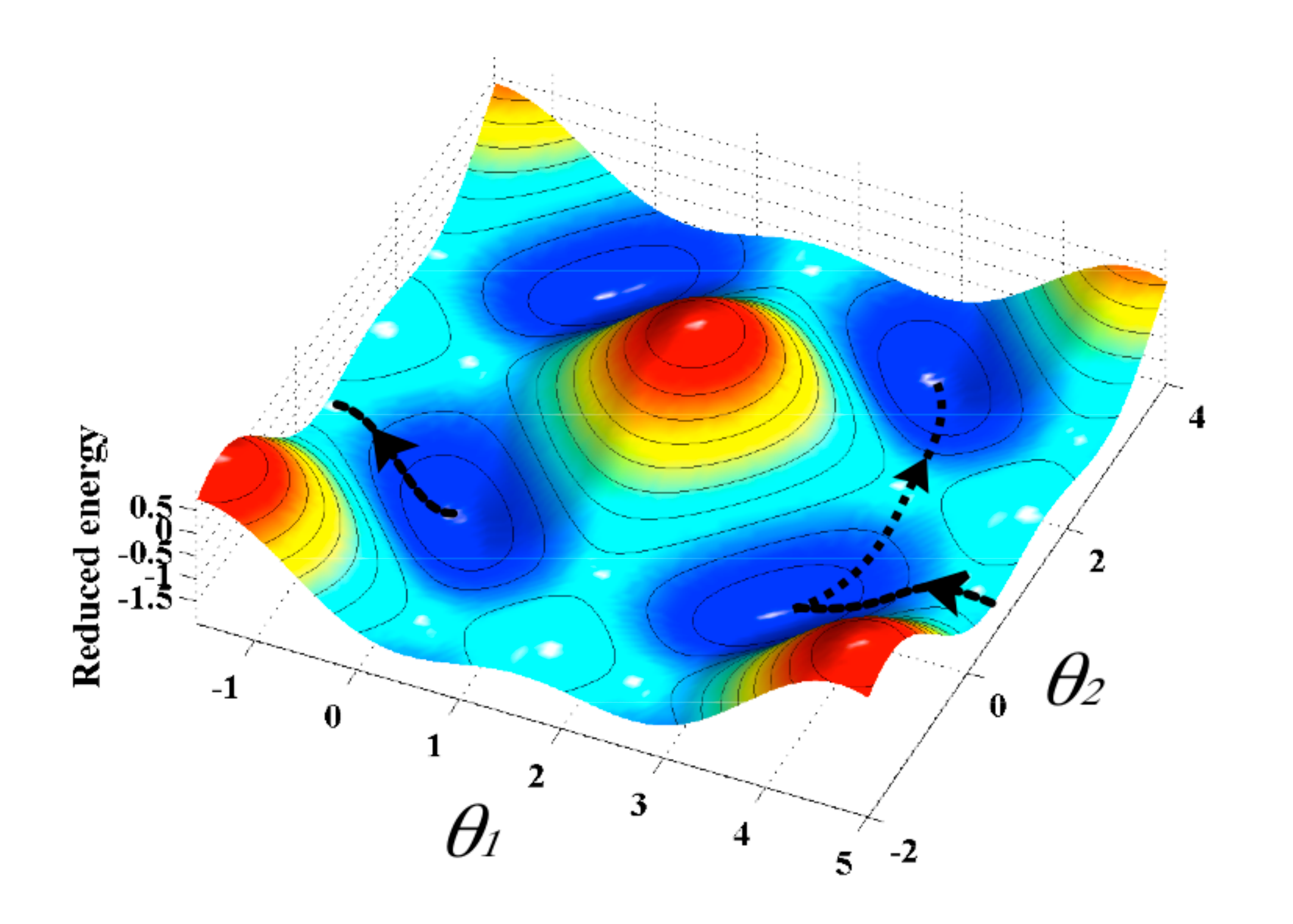}}
\subfloat[Strong Coupling $\delta=3$]{\includegraphics[width=7cm]{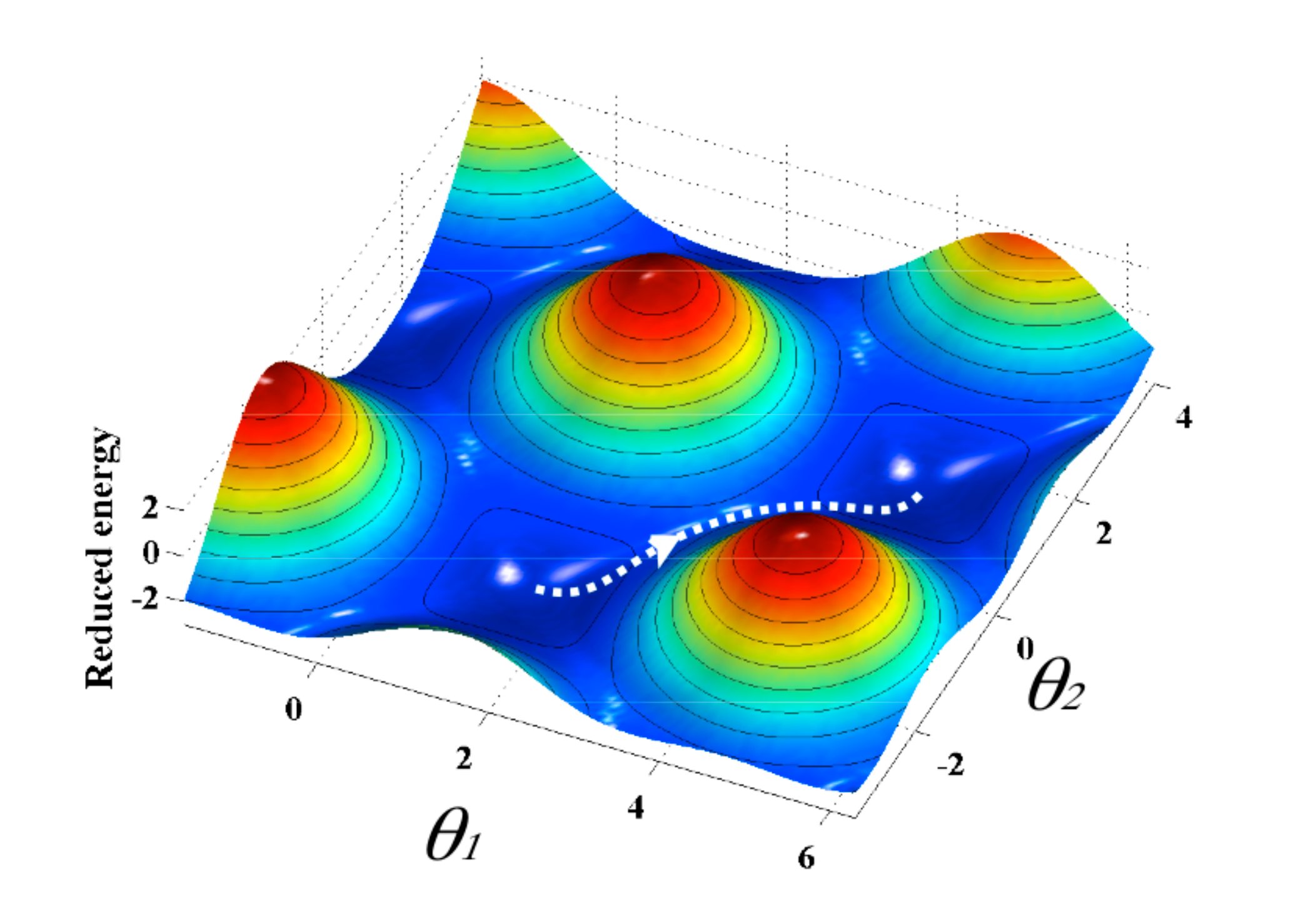}}
\caption{\label{fig:En-DMI-MD-LA}Energy potential surface of the different regimes for the DMI-MD with LA.}
\end{figure*}

The energy potential surface is shown in Fig. \ref{fig:En-DMI-MD-LA} (a). It can be seen that the net magnetic moment
goes from, say the minimum $(0,0)$ to the minimum $\left(\pi,\pi\right)$ through a two-step process that can proceed along two symmetrical
paths. Each one of these goes through the first saddle point $\left(-\frac{\pi}{2},\arcsin\left(\frac{d}{2k}\right)\right)$,
passes into the local minimum $\left(\pi,0\right)$ and crosses the saddle
point $\left(-\pi-\arcsin\left(\frac{\delta}{2k}\right),-\frac{\pi}{2}\right)$.

The switching rates corresponding to these two steps have the expression
\begin{eqnarray}
\Gamma^{\left(i\right)} & = & \left|\kappa_{i}\right|\sqrt{\frac{\sigma}{2\pi}}e^{-\frac{\sigma}{4}\left(4-\delta^{2}\right)},\quad i=1,2.\label{eq:RR-DMI-LAWC}
\end{eqnarray}
where the attempt frequencies $\kappa_{i}$ are computed numerically. 
Upon counting the symmetry factors, we obtain the total switching rate
\begin{eqnarray}
{\Gamma}^{DMI}_\mathrm{LAWC} & = & 2\left|\kappa\right|\sqrt{\frac{\sigma}{2\pi}}e^{-\frac{\sigma}{4}\left(4-\delta^{2}\right)}
\label{eq:TRR-MI-LAWC}.
\end{eqnarray}
\subsubsection{Strong Coupling}
In this regime, the DMI wins against the anisotropy field leading to a minimum with perpendicular
magnetic moments, lying in the plane normal to the anisotropy axes since the DMI vector $\mathbf{D}$
is oriented along the latter. As such, the minima are
\begin{equation*}
\left(\theta_{1},\theta_{2}\right)=\left(\pm\frac{\pi}{2},\mp\frac{\pi}{2}\right) 
\end{equation*}
while the states
\begin{equation*}
\left(\theta_{1},\theta_{2}\right)=\left(\pm\frac{\pi}{2},\pm\frac{\pi}{2}\right) 
\end{equation*}
are maxima and the saddle points now are 
\begin{equation}
\left(\theta_{1},\theta_{2}\right)=\left(0,0\right),\left(\pi,\pi\right),\left(0,\pi\right)\left(\pi,0\right). 
\end{equation}

Hence, the system may escape from the state $\left(-\frac{\pi}{2},\frac{\pi}{2}\right)$ into the
state $\left(\frac{\pi}{2},-\frac{\pi}{2}\right)$, thus reversing its resultant magnetic moment,
along two different paths comprising the saddle points $\left(0,0\right)\mbox{ and
}\left(\pi,\pi\right)$.
The switching rate of escape via one of these paths is given by
\begin{equation}
\Gamma^{DMI}_{\mathrm{LASC}} =
\alpha\left(\frac{2}{\pi}\right)^{3/2}\sqrt{\frac{\delta}{2\sigma}}\frac{\delta-2}{\delta+2}e^{
-\sigma\left(\delta-2\right)}.\label{eq:RR-DMI-LASC} 
\end{equation}

As is usually the case in the SC regime, the attempt frequency has been obtained analytically.
\begin{figure*}
\includegraphics[width=8cm]{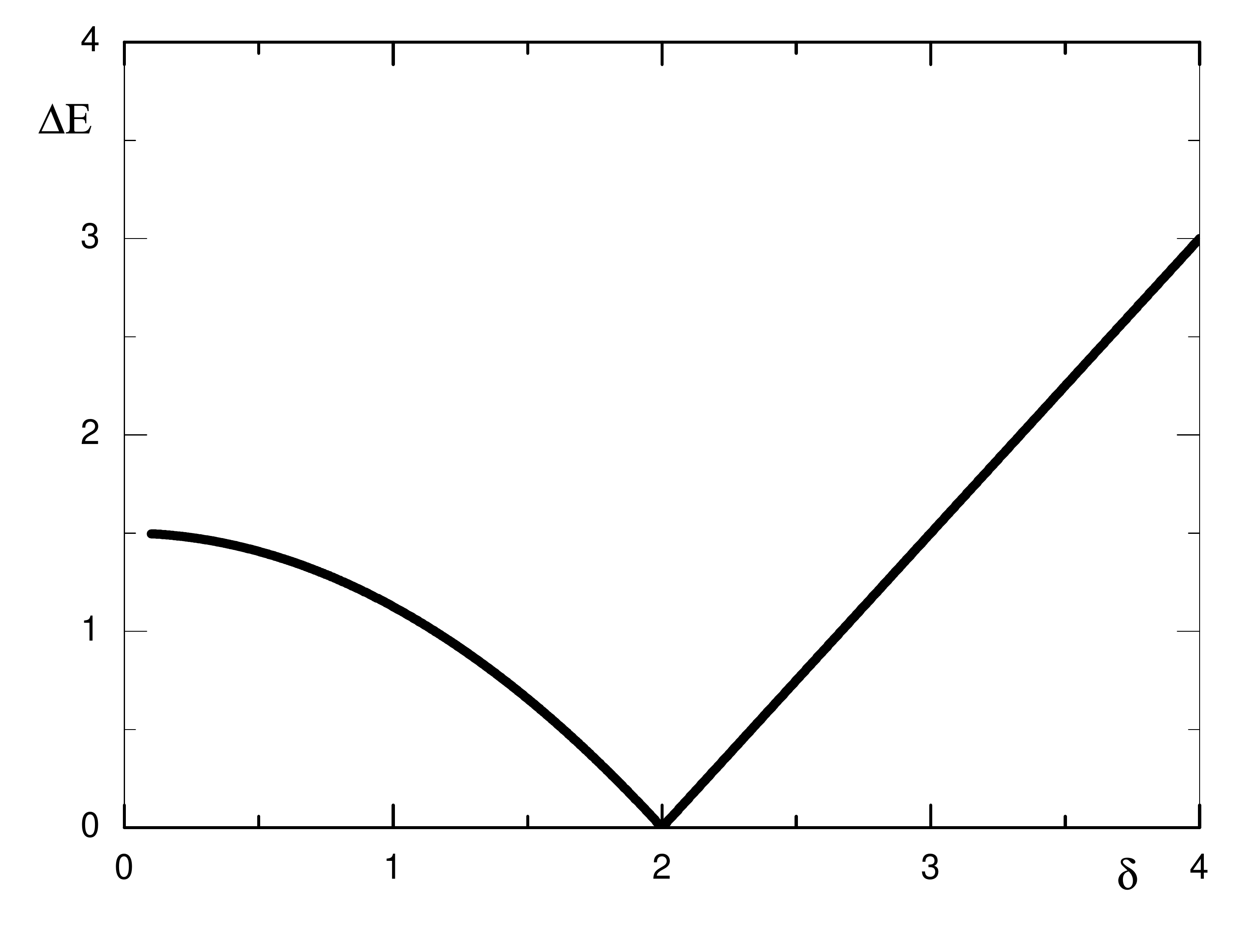}
\includegraphics[width=8cm]{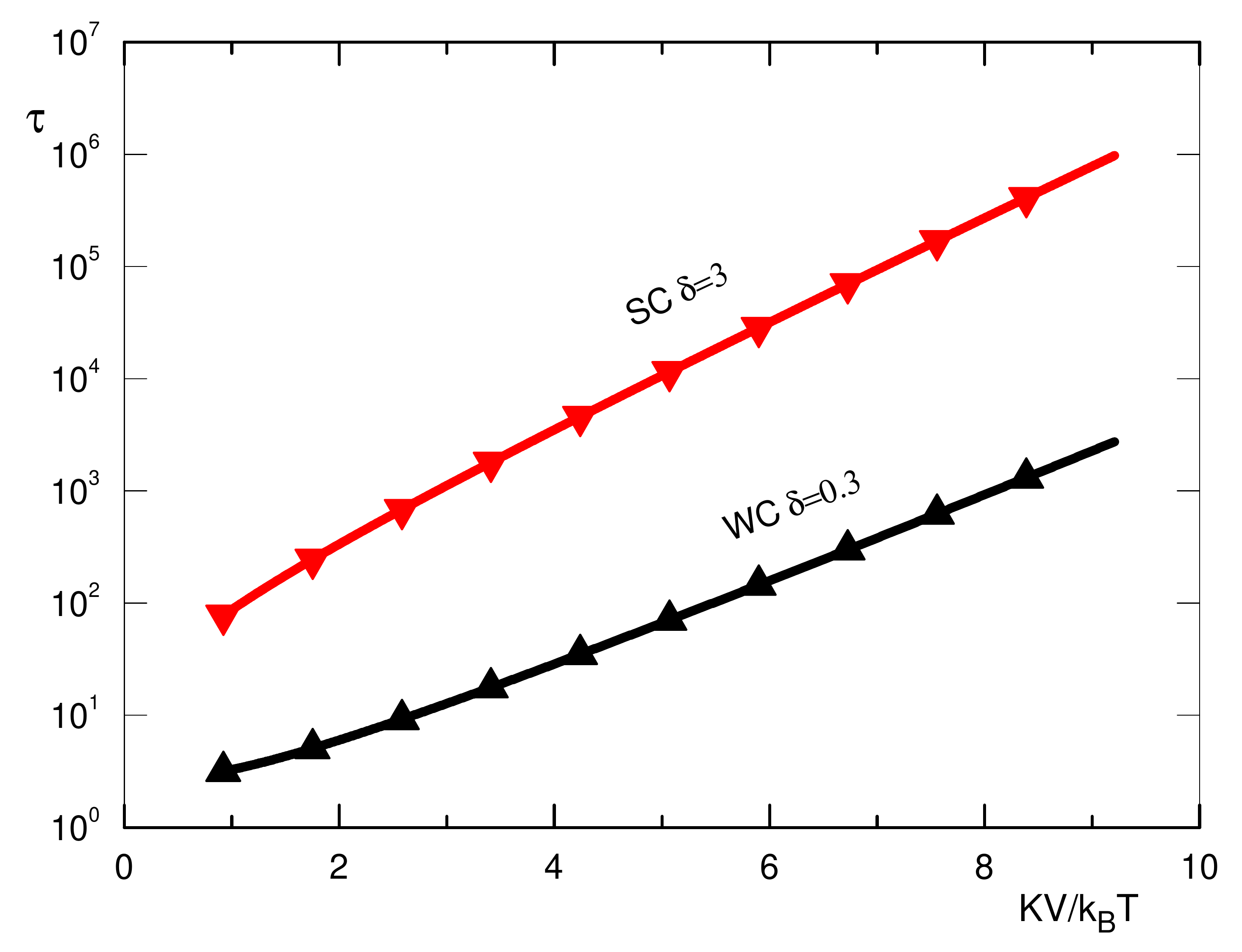}
\caption{\label{fig:RT-En-DMI-MD-LA}Energy Barrier (left) and reduced switching time (right) of the DMI magnetic dimer with LA for WC and SC.}
\end{figure*}

Fig. \ref{fig:En-DMI-MD-LA} shows the $3D$ energy potential surfaces for both WC and SC regimes. 
In Fig. \ref{fig:RT-En-DMI-MD-LA} we plot the energy barrier (left) and the
switching time (right) for the DMI-MD in the case of $\mathbf{D}$ parallel
to the anisotropy axes. It is seen that the energy barriers for the
two steps in the WC regime are equal and decrease quadratically with
the DMI strength $\delta$ [see the argument of the exponential
in Eq. (\ref{eq:TRR-MI-LAWC})]. At the critical value of the DMI coupling,
$\delta_{c}/k=2$, the energy barrier vanishes and immediately after
that it increases linearly with $\delta/k$, as can be seen in Eq. (\ref{eq:RR-DMI-LASC}). 

Here again we see that the stronger is the coupling the slower is the MD switching.
Note that in this regime, the saddle point corresponds to the state with the two magnetic moments
along the easy axes and, more importantly, parallel to the DM vector $\textbf{D}$. To go
through this saddle point the magnetic moments have to break free from the interaction and also to
circumvent the anisotropy due to the DM interaction.
This implies that the DMI leads to a longer switching time than the EI [see below]. Indeed, in the
latter case switching is achieved against the (exchange) coupling while in the former it is
achieved against the (DMI) coupling and the induced anisotropy.
\section{Most efficient coupling in a magnetic dimer}
In this section we present a pairwise comparison of the different
interactions with regard to their effect on the MD switching and on the
corresponding switching time. On one hand, we have the short-range
interactions EI and DMI, symmetric and anti-symmetric, respectively.
On the other hand, we have the long-range and antisymmetric interaction
DDI. We first compare the EI with DDI and investigate the effects pertaining to the MD bond. Then, comes the comparison
between the spins scalar-product and vector-product interactions,
\emph{i.e.} EI and DMI. Finally, we compare the DDI and DMI.

\subsection{EI versus DDI}

For the EI-MD, when the exchange coupling exceeds the critical value
the energy barrier {[}see Fig. 2 of Ref. \citealp{kac03epl-kac04jml}{]} becomes
independent of the exchange coupling as soon as the saturated ferromagnetic
state is reached. For the DDI-MD, the situation is fundamentally different
because the energy barrier continues to increase as the DDI coupling
increases, see \emph{e.g.} Eq. (\ref{eq:RR-LASC}) for the SC regime,
where $\Delta\mathcal{E}=\sigma\left(1+\xi\right)$. This is due to
the fact that the distance between the two magnetic moments, belonging
to the layers or to the magnetic nanoparticles, plays a crucial role.
Indeed, this distance cannot be smaller than a certain minimal value
that corresponds to the thickness of the nonmagnetic spacer (in the
case of two magnetic layers), or to the sum of radii of the two particles, or to the inter-atomic
distance. Therefore, it is understood that the $\xi$ axis must be cut off at a given value because 
the unlimited increase of $\xi$ simply reflects the unphysical asymptotic limit $d\rightarrow0$.

\begin{figure*}
\includegraphics[width=8cm]{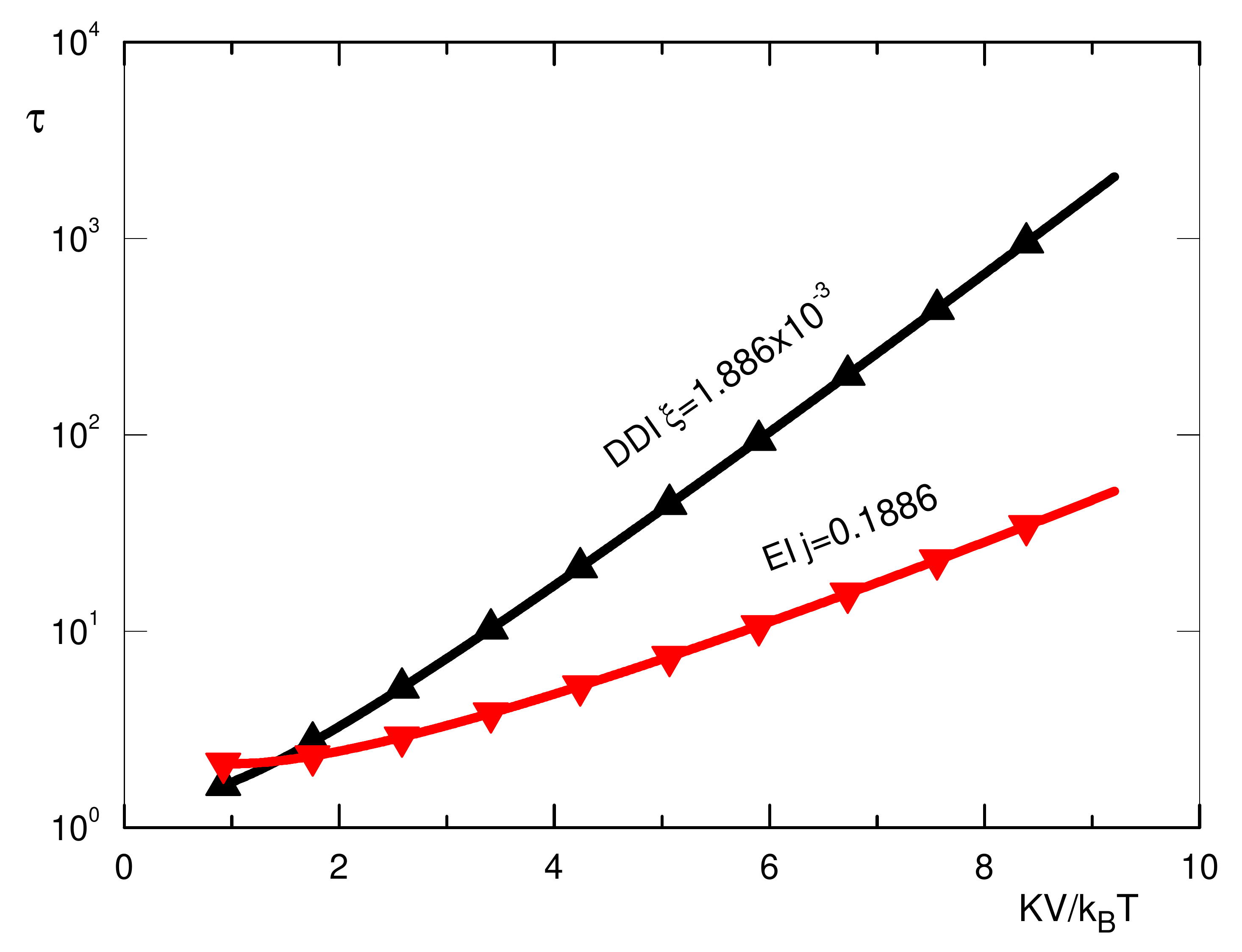}
\includegraphics[width=8cm]{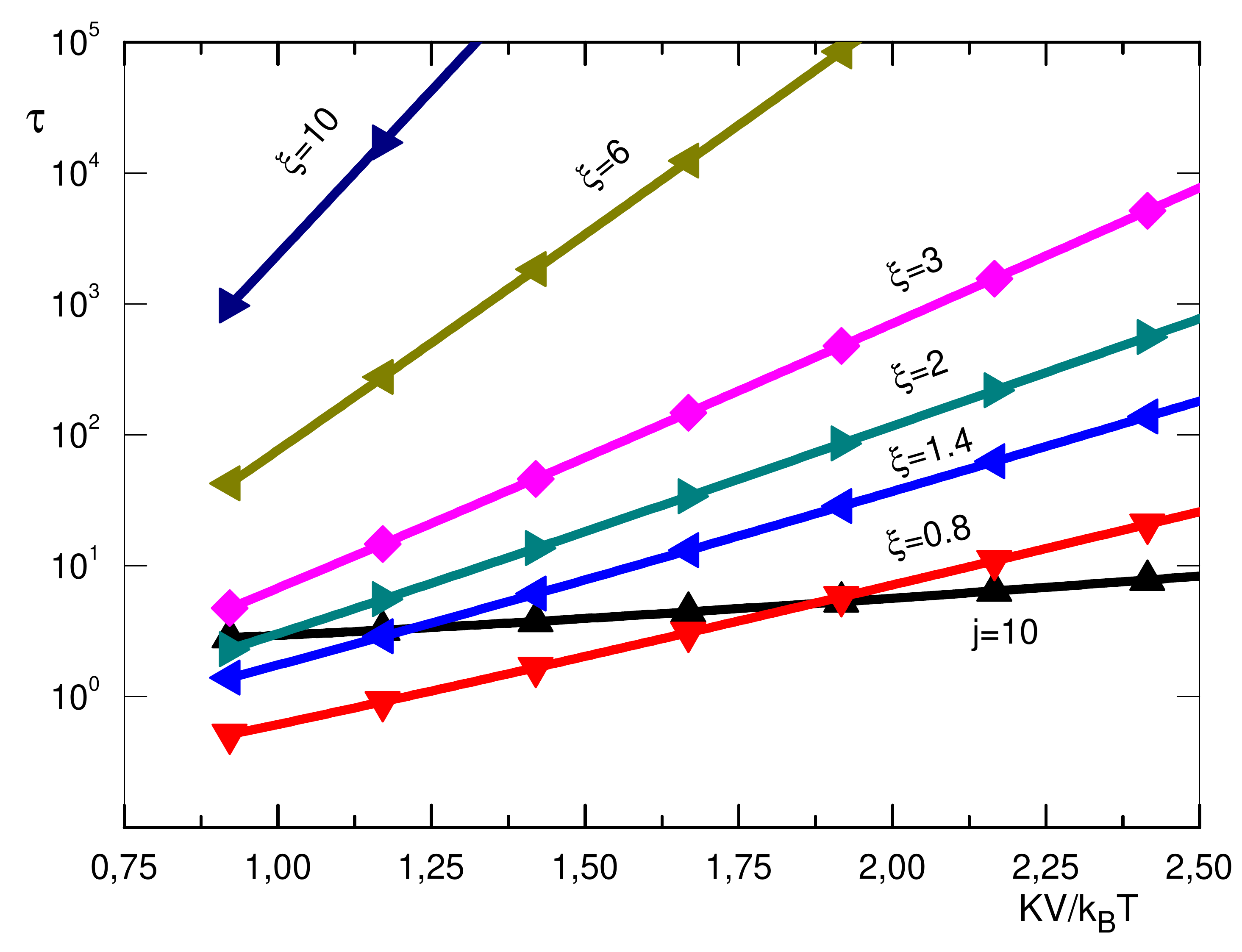}
\caption{\label{fig:RelaxRate-EIvsDDI}Reduced switching time, $\tau$, versus $\sigma=KV/k_{B}T$ for the EI- and DDI-MD, in the absence
of the magnetic field, for weak-coupling regime (left) and strong-coupling regime (right).}
\end{figure*}

Let us now compare the switching times of the MD with LA when coupled via
EI or DDI, in both the WC and SC regimes. The results are shown in
Fig. \ref{fig:RelaxRate-EIvsDDI} where the (reduced) switching time
is plotted as a function of $\sigma=KV/k_{\mathrm{B}}T$.
For these calculations, both the EI-MD and DDI-MD switch from the
same initial state $(0,0)$ into the same final state $(\pi,\pi)$.

As was discussed earlier, apart from the fact that the switching
time obviously increases with $\sigma$ (or with decreasing temperature)
for both EI-MD and DDI-MD, we see that for both coupling regimes there
is a critical value $\sigma_{c}$ at which the switching times corresponding
to EI and DDI intersect each other. 
Indeed, the DDI is always faster than the EI for low values of $\sigma$ (below $\sigma_{c}$)
and the situation reverses for values of $\sigma$ higher than $\sigma_{c}$.
The EI energy barrier is constant while that of DDI continues to grow. So, below $\sigma_{c}$ the 
prefactor of the switching time prevails and the DDI is more favorable for
a fast switching. However, as $\sigma_{c}$ is exceeded, the energy
barrier prevails over the prefactor and thereby the ever growing DDI
energy barrier leads to a slower switching than via EI. The expression
of $\sigma_{c}$ is obtained in terms of the ratio of the
switching rates in Eqs. (\ref{eq:Gamma_tsp_j>jc}) and (\ref{eq:RR-LASC}), which is of
the form $\mathrm{Prefactor}/e^{\sigma\xi}$ with\[
\mathrm{Prefactor}=\frac{\kappa}{\sqrt{2}}\sqrt{\frac{j\left(j-1\right)\left(\xi+2\right)}{\xi\left(3\xi-2\right)}}\frac{3\xi+2}{j+1}\]
where 

\begin{eqnarray*}
\kappa & = & 1-\frac{\xi}{2}-\sqrt{\left(\frac{3}{2}\xi+1\right)^{2}+\frac{2}{\alpha^{2}}\xi\left(12+\xi\right)}.
\end{eqnarray*}

More precisely, we have

\begin{eqnarray*}
\sigma_{c}\left(j,\xi,\alpha\right) & = & \frac{1}{\xi+1}\ln\left(\mathrm{Prefactor}\right).
\end{eqnarray*}

The critical value $\sigma_{c}$ is a decreasing function of the ratio
$\xi/j$, which is simply due to the fact that the stronger the DDI
the smaller is $\sigma_{c}$ at which the energy barrier prevails
over the prefactor of the switching time.

In conclusion, at low temperature, the EI-MD switches faster than
DDI-MD.

\subsection{EI versus DMI}

\begin{figure}
 \includegraphics[width=8cm]{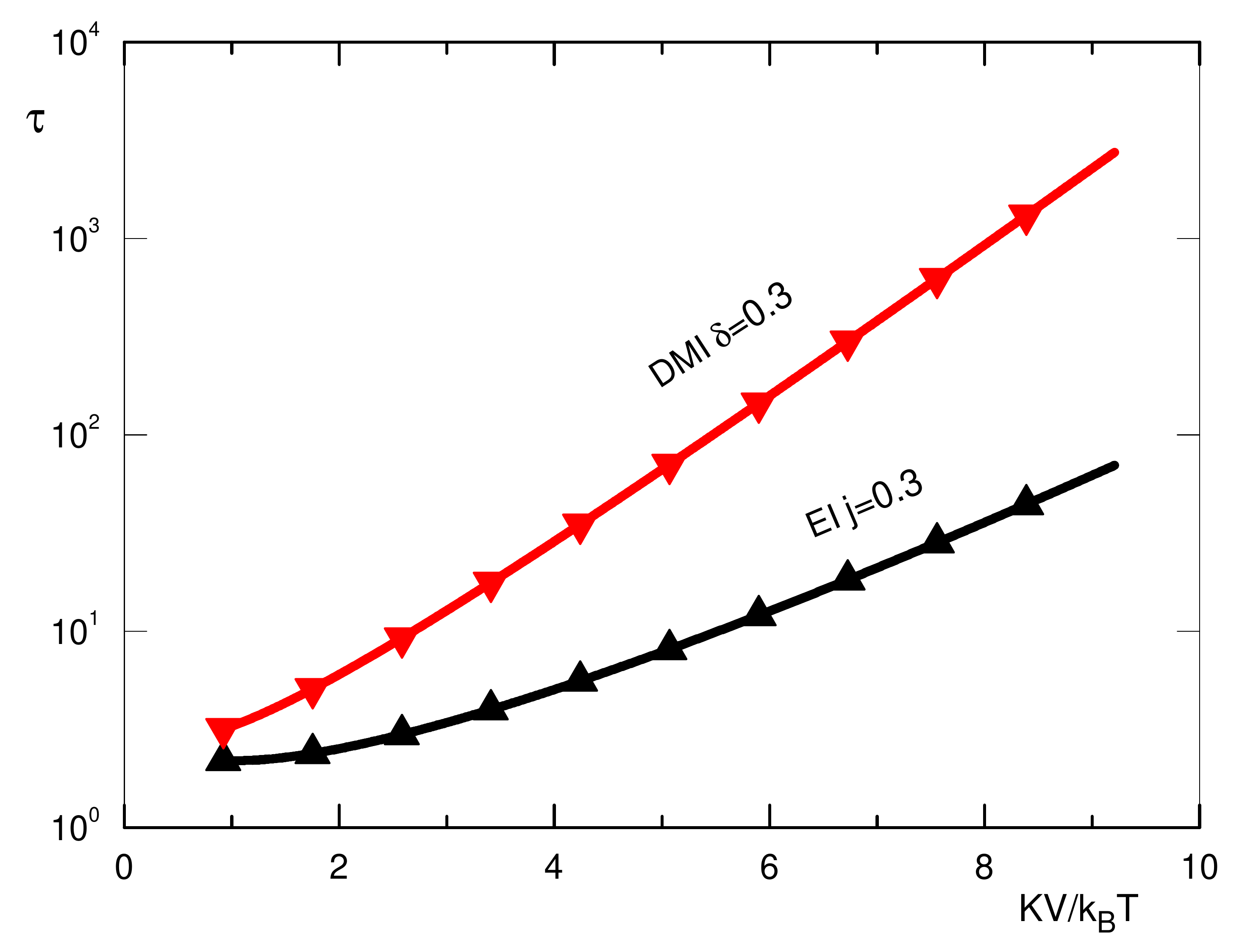}
 \caption{\label{fig:EIvsDMI_WC}Reduced switching time, $\tau$, versus $\sigma=KV/k_{B}T$ for the EI and DMI-MD, in the absence
of the magnetic field, for weak-coupling regime.}
\end{figure}

Fig. \ref{fig:EIvsDMI_WC} shows a comparison between the reduced 
switching times of the EI-MD and the DMI-MD in the weak coupling regime.

The initial and final states are identical for both interactions and are given by
 $\left(\theta_{1},\theta_{2}\right)=\left(\pi,\pi\right)$ (initial state), 
 $\left(\theta_{1},\theta_{2}\right)=\left(0,0\right)$ (final state).
 
Unlike the previous cases (Fig. \ref{fig:RelaxRate-EIvsDDI}) there is no intersection of the two
curves at finite values of $\sigma$ in a range where our approach is applicable, \emph{i.e.} where
the second-order expansion of the energy is applicable.

From these results we see that the EI-MD has a shorter switching time than the DMI-MD. Indeed, in
the weak coupling regime, for the dimer to switch one of its magnetic moments has to cross a saddle
point into an intermediate state. In so doing, it has to circumvent the energy barrier associated
with its coupling to the second moment.
In the case of the DM coupling, in addition to the breaking free from the coupling there is a
constraint related with the orientation imposed by the DM vector D, \emph{i.e.} the inherent
anisotropy.

\subsection{DDI versus DMI}
\begin{figure}
 \includegraphics[width=8cm]{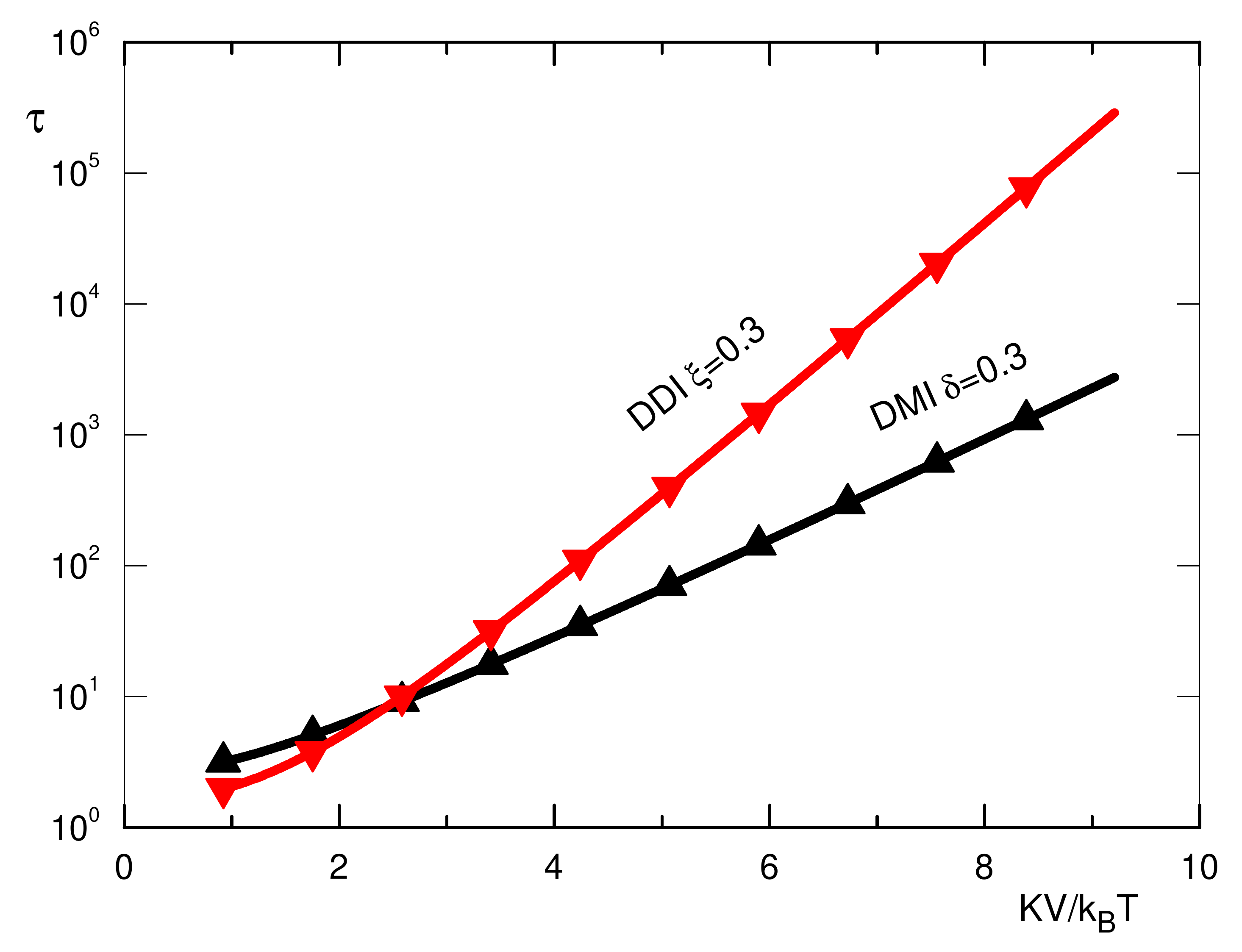}
  \caption{\label{fig:DDIvsDMI_WC}Reduced switching time, $\tau$, versus $\sigma=KV/k_{B}T$ for the
DDI-MD and DMI-MD, in the absence of the magnetic field, for weak-coupling regime.}
\end{figure}

In Fig. \ref{fig:DDIvsDMI_WC}, we compare the reduced switching times of the DDI-MD and the DMI-MD in
the weak coupling regime. Two different curves for each interaction are presented.

The initial and final state are identical for both interactions and are given by
 $\left(\theta_{1},\theta_{2}\right)=\left(\pi,\pi\right)$ (initial state), 
 $\left(\theta_{1},\theta_{2}\right)=\left(0,0\right)$ (final state).
 
We first see that there exists a $\sigma_{c}$ and that DDI leads to a faster switching than the DMI for $\sigma<\sigma_{c}$.

Switching in both cases is performed against the spin coupling and the anisotropy. However, the
latter has a stronger effect in the case of the DMI dimer.
\section{Conclusion}
We have considered a magnetic dimer as a model system of two magnetic moments, atomic or macroscopic, coupled by either exchange, dipole-dipole, or Dzyalozhinski-Moriya interaction.
This is a quite general system since the two magnetic moments may be those of two thin layers coupled by an effective interaction through a non magnetic spacer, two magnetic nanoparticles in a hosting matrix or on a substrate, or still two atomic magnetic moments.
We have identified various coupling regimes and investigated the switching mechanisms of the system in each regime and in different anisotropy configurations. In each situation, we have computed the energy barrier and, for the high-to-intermediate damping, we used Langer's approach to compute the switching rate and, in some cases, provided the corresponding analytical expressions. 

We have investigated how the energy barriers are affected by the coupling. For instance, for the
dipole-dipole interaction we find that the energy barrier may either increase or decrease with the
coupling depending on the coupling regime. In the weak-coupling regime, we find that the
switching rate, as a function of temperature, does not follow the simple Arrhenius law because the
prefactor dominates over the exponential.
Furthermore, transverse anisotropy or equivalently, rather thin magnetic films, seem to exhibit the
fastest switching process, as compared with the longitudinal or mixed anisotropy.

We then compared the three interactions with regard to their efficiency in switching the magnetic
dimer. Comparing exchange and DDI led to the conclusion that below some critical temperature the
exchange-coupled MD switches faster than the dipolar-coupled MD.
Next, comparing the isotropic and anisotropic exchange and Dzyaloshinskii-Moriya interactions we
have seen that in the latter case the inherent anisotropy makes the switching longer.
Altogether, we have
\begin{equation*}
\tau_\mathrm{EI} < \tau_\mathrm{DMI} < \tau_\mathrm{DDI}
\end{equation*}
which is compatible with the fact that the corresponding couplings are ordered in the following way
\begin{equation*}
\lambda_\mathrm{EI} > \lambda_\mathrm{DMI} > \lambda_\mathrm{DDI}
\end{equation*}
and that $\tau\sim 1/\lambda$.

In a pure material, \emph{i.e.} without too many impurities, it turns out that the fastest
recovering of the magnetic state and thereby that of the system magnetization occurs via the
exchange coupling. In this work, we provide details of how this switching occurs.

We have already started a few experiments for investigating the dynamics of coupled thin films grown by our collaborators. We intend to perform various measurements of FMR with varying field and frequency using a network analyzer. In addition, the slow dynamics of the dimer may be probed by measuring the isothermal and thermoremanent magnetization by a commercial SQUID in a wide range of temperature.

For a closer comparison with experiments we need to consider more general situations with
arbitrary directions of the two anisotropy axes in an oblique magnetic field. Such calculations
will be performed numerically in a subsequent work.

\bibliography{/usr/share/texmf-texlive/bibtex/bib/hk/hkbib}

\appendix
\section{Details for WC of DDI-MD with MA\label{apx:WC-DDI-MA}}

The stationary states are given by

\begin{eqnarray}
\theta_{1},\theta_{2} & = & \frac{\pm_{a}\arccos x_{+}\pm_{b}\arccos x_{-}}{2}\nonumber\\
\theta_{1},\theta_{2} & = & \pm_{c}\pi+\frac{\pm_{d}\arccos x_{+}\pm_{e}\arccos x_{-}}{2}\label{eq:SS-WC-DDI-MA}
\end{eqnarray}
where 
\begin{eqnarray*}
x_{+}^{+} & = & \pm1,\quad x_{-}^{+}=\pm1,\\
x_{+}^{-} & = & \pm\sqrt{\frac{36\xi^{2}+9\xi^{4}}{36\xi^{2}+16}},x_{-}^{-}=\pm\sqrt{\frac{9\xi^{4}+4\xi^{2}}{16+4\xi^{2}}}.
\end{eqnarray*}
and the subindex $a...e$ indicates independence between different $\pm,\mp$ signs.

In the weak coupling, the minima of the MD are located at

\begin{eqnarray*}
\theta_{1} & = & \pm\frac{\arccos x_{+}^{-}-\arccos x_{-}^{-}}{2},\\
\theta_{2} & = & \pm\frac{\arccos x_{+}^{-}+\arccos x_{-}^{-}}{2}
\end{eqnarray*}
and
\begin{eqnarray*}
\theta_{1} & = & \mp\pi\mp\frac{\arccos x_{+}^{-}-\arccos x_{-}^{-}}{2},\\
\theta_{2} & = & \pm\pi\mp\frac{\arccos x_{+}^{-}+\arccos x_{-}^{-}}{2}
\end{eqnarray*}
where the sign of $x_{\pm}^{-}$ is taken as positive. The maxima are at
\begin{eqnarray*}
\theta_{1} & = & \pm\frac{\arccos x_{+}^{-}-\arccos x_{-}^{-}}{2},\\
\theta_{2} & = & \pm\frac{\arccos x_{+}^{-}+\arccos x_{-}^{-}}{2}
\end{eqnarray*}
and 
\begin{eqnarray*}
\theta_{1} & = & \pm\pi\mp\frac{\arccos x_{+}^{-}+\arccos x_{-}^{-}}{2},\\
\theta_{2} & = & \pm\pi\mp\frac{\arccos x_{+}^{-}-\arccos x_{-}^{-}}{2}
\end{eqnarray*}
where the sign of $x_{\pm}^{-}$ is taken as negative. Hence, the saddle point are at
\begin{eqnarray*}
\left(\theta_{1},\theta_{2}\right) & = & \left(0,\pm\pi\right),\left(\pm\pi,0\right),\left(\pm\pi,\pm\pi\right),\\
 &  & \left(\pm\pi,\mp\pi\right),\left(\pm\frac{\pi}{2},\pm\frac{\pi}{2}\right),\left(\pm\frac{\pi}{2},\mp\frac{\pi}{2}\right).
\end{eqnarray*}

The switching rates for the first and second step in the WC regime read

\begin{widetext}
\begin{eqnarray}
\Gamma_{\mathrm{MAWC}}^{(1)} & = & \frac{\left|\kappa\right|}{2\pi}\sqrt{\frac{W_{+}^{1}W_{-}^{1}}{\left(\xi+2r\right)\left|\xi-2r\right|\left(1+\xi+r\right)}}\sqrt{\frac{V_{+}^{1}V_{-}^{1}}{\left(1+\xi-r\right)}}e^{\sigma+\sigma\xi+\varepsilon_{m}^{1}(0)}\label{eq:RR-MAWC},\\
\Gamma_{\mathrm{MAWC}}^{(2)} & = & \frac{\left|\kappa\right|}{4\pi}\sqrt{\frac{W_{+}^{2}W_{-}^{2}}{\left(2\xi+r_{p}\right)\left(1+2\xi+r\right)}}\sqrt{\frac{V_{+}^{2}V_{-}^{2}}{\left|2\xi-r_{p}\right|\left(1+2\xi-r_{p}\right)}}e^{\sigma+2\sigma\xi+\varepsilon_{m}^{2}(0)},\nonumber 
\end{eqnarray}
where
\begin{equation*}
\varepsilon_{m}^{j}(0)=-\sigma\left(1+\left(C_{p1}^{j}\right)^{2}-\left(C_{p2}^{j}\right)^{2}\right)-\sigma\xi\left(2C_{p1}^{j}C_{p2}^{j}+Q_{p}^{j}\right),
\end{equation*}

\begin{equation*}
V_{\pm}^{j}=N_{V}^{j}\pm\frac{1}{2}R_{P}^{j},\quad W_{\pm}^{j}=N_{t}^{j}\pm R_{t}^{j}, 
\end{equation*}

\begin{equation*}
N_{t}^{j}=\left[\left(C_{p1}^{j}\right)^{2}-\left(C_{p2}^{j}\right)^{2}\right]+\xi C_{p1}^{j}C_{p2}^{j}+\frac{\xi}{2}Q_{p}^{j},\quad N_{V}^{j}=-\frac{1}{2}\left(1-\left(C_{p2}^{j}\right)^{2}\right)-\frac{\xi}{2}Q_{p}^{j},
\end{equation*}

\begin{eqnarray*}
R_{t}^{j} & = & \sqrt{\left(1-\left(C_{p1}^{j}\right)^{2}-\left(C_{p2}^{j}\right)^{2}\right)^{2}+\frac{\xi^{2}}{4}\left(C_{p1}^{j}C_{p2}^{j}+2Q_{p}^{j}\right)^{2}},\\
R_{P}^{j} & = & \sqrt{\left(1-\left(C_{p2}^{j}\right)^{2}\right)^{2}+\xi^{2}\left(Q_{p}^{j}\right)^{2}},\quad Q_{p}^{j}\equiv\sqrt{\left(1-\left(C_{p1}^{j}\right)^{2}\right)\left(1-\left(C_{p2}^{j}\right)^{2}\right)},
\end{eqnarray*}
\end{widetext}
with
\begin{eqnarray*}
r & = & \sqrt{\xi^{2}+1},\quad r_{p}=\sqrt{\xi^{2}+4},\\
C_{pi}^{j} & \equiv & \cos\theta_{i}^{j}.
\end{eqnarray*}
$i=1,2$ refers to the $i^{\mathrm{th}}$ layer and $j=1,2$ refers
to the $j^{\mathrm{th}}$ minimum.

\end{document}